\documentclass[aps,prb,reprint,showpacs]{revtex4-1}
\usepackage{epsfig}\usepackage{subfigure}
\usepackage{dcolumn}
\usepackage{graphicx,graphics}
\usepackage{amsmath}
\usepackage{amssymb}
\usepackage{bm}
\usepackage{color}
\usepackage[utf8]{inputenc}
\newcommand{\be}{\begin{equation}}
\newcommand{\ee}{\end{equation}}
\newcommand{\bea}{\begin{eqnarray}}
\newcommand{\eea}{\end{eqnarray}}
\newcommand{\bse}{\begin{subequations}}
\newcommand{\ese}{\end{subequations}}

\begin{document}

\author{Mahmoud M. Asmar}
\email{asmar@lsu.edu} \affiliation{Department of Physics and Astronomy, Louisiana State University, Baton Rouge, LA 70803-4001 }
\author{Daniel E. Sheehy }
\email{sheehy@lsu.edu} \affiliation{Department of Physics and Astronomy, Louisiana State University, Baton Rouge, LA 70803-4001 }
\author{Ilya Vekhter}
\email{vekhter@lsu.edu}\affiliation{Department of Physics and Astronomy, Louisiana State University, Baton Rouge, LA 70803-4001 }

\title{Topological phases of topological insulator thin films}

\begin{abstract}
We study the properties of a thin film of topological insulator material. We treat the coupling between helical states at opposite surfaces of the film in the properly-adapted tunneling approximation, and show that the tunneling matrix element oscillates as function of both the film thickness and the momentum in the plane of the film for Bi$_2$Se$_3$ and Bi$_2$Te$_3$. As a result, while the magnitude of the matrix element at the center of the surface Brillouin Zone gives the gap in the energy spectrum, the sign of the matrix element uniquely determines the topological properties of the film, as demonstrated by explicitly computing the pseudospin textures and the Chern number. We find a sequence of transitions between topological and non-topological phases, separated by semimetallic states, as the film thickness varies. In the topological phase the edge states of the film always exist but only carry a spin current if the edge potentials break particle-hole symmetry. The edge states decay very slowly away from the boundary in Bi$_2$Se$_3$, making Bi$_{2}$Te$_{3}$, where this scale is shorter, a more promising candidate for the observation of these states.  Our results hold for free-standing films as well as heterostructures with large-gap insulators.

\end{abstract}

\date\today

\maketitle
\section{Introduction}

The theoretical prediction and experimental discovery of  two dimensional (2D) and three dimensional (3D) topological insulators (TIs)~\cite{kane,BHZmodel,bernevig,TIhamiltonian,Rev1,Rev2,ARPES1,ARPES2,ARPES3,ARPES4,ARPES5,ARPES6} has led to a strong effort aimed at understanding and utilizing their unique electronic properties.  While electronically insulating (gapped) in the bulk, TIs possess gapless states at their boundaries. In the 3D bulk compounds the presence of topological surface states described by an effective 2D massless  Dirac-like Hamiltonian is confirmed by angle-resolved photoemission spectroscopy (ARPES) experiments and other measurements on materials such as Bi$_{2}$Se$_{3}$, Bi$_{2}$Te$_{3}$ and Sb$_{2}$Te$_{3}$ ~\cite{ARPES1,ARPES2,ARPES3,ARPES4,ARPES5,ARPES6}.

Many modern applications integrate thin films and  small-size components, and TIs are no exception~\cite{bulk2,assymetry,nanorib}. An important point is that the TI surface states are characterized by a length scale over which their wave functions decay into the bulk. In materials such as Bi$_2$Se$_3$ this length is on the order of several nanometers.  Consequently, when the film is sufficiently thin, the gapless TI surface states on opposite surfaces hybridize, leading to a gap in the spectrum~\cite{assymetry,thingap1,thingap2,thingap3}. Refs.~\onlinecite{thingap3,thingap2} showed that for Bi$_2$Se$_3$ this gap has a non-monotonic dependence on the film thickness, and argued that the thickness also changes the topological properties of the resulting hybridized states. They found two types of transitions between trivial and non-trivial gapped topological phases: one where the gap closed at the transition point, and another where the gap remained finite. The latter result seemingly contradicts the established theory of topological phase transitions in non-interacting systems~\cite{TKNN,Nagaosa} that requires a gapless state to appear at the point where the corresponding topological quantum number changes. Hence on general grounds we expect a Dirac semimetal to appear at such phase transitions~\cite{diracsemimetal1}. This apparent contradiction motivated us to revisit the study of the topological phases in TI thin films.

In this paper we consider a flat free standing thin film of a topological insulator material within the tunneling formalism, i.e. assuming weak hybridization. This is justified because the decay scale of the surface state in the best studied TIs, such as Bi$_{2}$Se$_{3}$ and Bi$_{2}$Te$_{3}$, is comparable to the size of the quintuple layer (QL), the basic structural unit of these materials. We demonstrate that this method requires careful consideration of the behavior of the wave function at the film boundaries.  Consequently we first develop a general tunneling approach valid for heterostructures where the TI film is sandwiched between other, topologically trivial, semiconductors or insulators, and then apply it to the problem of a free standing film. This allows us to elucidate the relevant physics and the origin of the hybridization. We compute the tunneling matrix elements between surface states,  derive the effective Hamiltonian for the film, and determine its energy spectrum and topological properties. The crucial part of our analysis that was missing in previous work is accounting, non-perturbatively, for the dependence of the decay length (and, consequently, the tunneling matrix elements) on the momentum in the plane of the film, ${\bm k}$. This dependence controls the band dispersion away from the zone center, and is necessary to determine the topological character of the carriers~\cite{bernevig}.

While the direct spectral gap at ${\bm k}=0$ (the ``mass'' term) agrees with the results of Refs.~\onlinecite{thingap3,thingap2}, the properties at finite ${\bm k}\neq 0$, such as the energy spectrum, and especially topological properties that depend on the band curvature, differ from those obtained via perturbative inclusion of $k^2$ terms in previous work. In particular, we show that the perturbative method leads to spurious topological phase transitions that are absent in our formalism.

We show that the topological phases of the thin film and the associated pseudo-spin windings in momentum space are uniquely determined by the sign of the tunneling matrix element at zero momentum, and confirm it by an explicit numerical calculation of the Chern number. Low energy massless and linearly-dispersing edge states appear at the sample boundaries in the topological phase. These edge states preserve time reversal symmetry (TRS), and, while they carry a pseudo-spin current, the physical spin associated with these currents vanishes unless the edges of the sample break particle-hole symmetry. We predict the experimental signatures of the edge states to be very weak in the Bi$_2$Se$_3$ films, but stronger in Bi$_2$Te$_3$. Finally, consistent with our expectations, we always find linearly dispersing gapless states at the boundary between the topological and trivial phases.

 The remainder of the paper is organized as follows.
 In Sec. \ref{tunnelingsec} we develop the tunneling formalism for a general junction involving TIs and non-TI materials,  and adapt it to the free-standing TI thin film. Recognizing the need to include the ${\bm k}$ dependence beyond the leading order expansion in small ${\bm k}$, we  revisit the problem of a single TI-Insulator junction and give the relevant solution for the interface states in Sec.~\ref{interfacestatessec}. In Sec.~\ref{effectivehamiltoniansec} we obtain the low energy effective Hamiltonian and the general low-energy band structure for a free standing TI thin film.  Since the results depend sensitively on the parameters of the Hamiltonian describing the bulk, in Sec. \ref{spectrumsec} we discuss the dependence of the spectrum and the gap on the choice of specific TI material and the film thickness.  To gain insight into the topological properties of the film, in Sec.~\ref{massparamiter}, we analyze the pseudo-spin textures associated with the band structure, and show when it is non-trivial. We complement this analysis by computing the Chern number for the thin film in Sec.~\ref{chernumbersec}, and verify the non trivial topological character of the film by demonstrating the existence of the edge states in Sec.~\ref{edgestatesec}.  In Sec.~\ref{conclusions} we provide brief concluding remarks.

\section{Tunneling approach and the effective Hamiltonian}
  Gapless surface states at the boundary between a TI and vacuum (or a wide gap insulator) are confined within a length $\lambda^{-1}$ of the  surface. Consequently, for TI films with thicknesses comparable to the surface state decay length $ \lambda^{-1}$, the states at opposite surfaces hybridize, and a spectral gap opens. In this section we describe the tunneling formalism and select the basis wave functions that we use to find the  parameters of the low energy effective Hamiltonian for these hybridized surface states.

\subsection{General tunneling formalism}\label{tunnelingsec}
\begin{figure}
  \centering
  \includegraphics[scale=0.25]{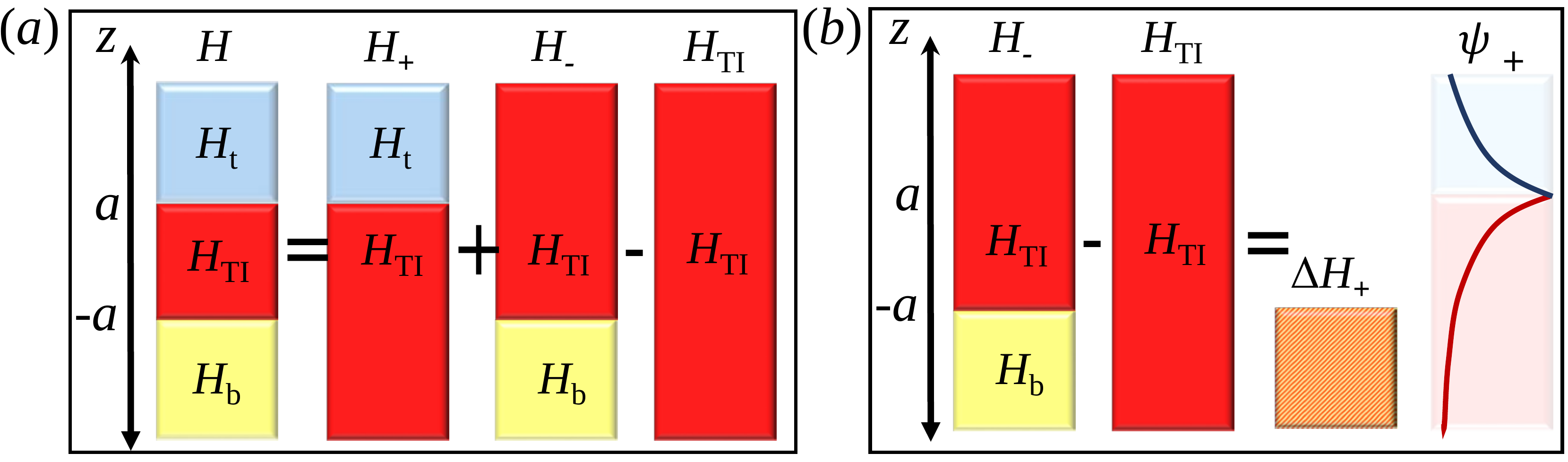}
  \caption{Schematic representation of the heterojunction with a TI of thickness $2a$ between semi-infinite top (${\rm I}_{t}$) and bottom (${\rm I}_{b}$) topologically trivial insulators. (a) Tunneling decomposition of the junction, described by Eq.~\eqref{exacthamiltonian}, as the sum of the two interfaces (${\rm I}_{t}$-TI and TI-${\rm I}_{b}$), and the coupling term, see text and Eq.~\eqref{decompositiion}. (b) Since the perturbing Hamiltonian for the eigenstate of the top (``$+$'') interface is only non-zero in the space below the bottom interface, exponential tails of the wave function are critical, see Eq.~\eqref{DeltaH_gen}.}\label{fig1}
\end{figure}

In the tunneling approximation the wave function of the low-energy state in a thin film is written as a linear combination of the wave functions of the metallic states that would exist at a single interface at the top and the bottom of the film respectively. The approach is essentially equivalent to the well known method of linear combination of atomic orbitals (LCAO) in quantum chemistry, and is illustrated in Fig.~\ref{fig1}. The key step is  the identification of the perturbing Hamiltonian that couples the two interface states. As we demonstrate below, this coupling becomes ill-defined in the limit of a vacuum termination. We therefore use a regularization procedure whereby we first solve the more general problem of a thin film of a TI material sandwiched between two topologically trivial semiconductors or insulators, and subsequently set the energy gap in the latter to infinity to model a vacuum termination.
 An important advantage of this approach is that it can also describe a film on a substrate as well as a real  I-TI-I junction, including interface potentials~\cite{interfacesymmery}, but in this work our focus is on the free-standing thin film.

To this end we consider the setup shown in Fig.~\ref{fig1} (a), described by the Hamiltonian
\begin{eqnarray}\label{exacthamiltonian}
H=&H_{\rm TI}&\Theta(z+a)\Theta(a-z) \nonumber\\ &+&H_{{\rm t}}\Theta(z-a)+H_{{\rm b}}\Theta(-a-z)\,.
\end{eqnarray}Here $\Theta(z)$ is the Heaviside step function, $H_{{\rm TI}}$ describes the TI, while $H_{{\rm t}}$ and $H_{{\rm b}}$ describe the top and bottom insulators respectively. We take the Hamiltonian for the TI in the form of the low-energy ${\bm k}\cdot{\bm p}$ approximation near the $\Gamma$ point derived by Liu, \textit{et al.}~\cite{TIhamiltonian} for materials such as Bi$_2$Se$_3$, Bi$_2$Te$_3$ and Sb$_2$Te$_3$. In the parity $(+,-)$ and spin $(\uparrow,\downarrow)$ basis,  $\psi^{T}=(\psi_{+\uparrow},\psi_{-\uparrow},\psi_{+\downarrow},\psi_{-\downarrow})$, the Hamiltonian $H_{{\rm TI}}$ reads~\cite{TIhamiltonian},
\begin{equation}\label{TIH}
H_{{\rm TI}} = \sigma_{0}\tau_{z}(M-B_{1}k_{z}^2-B_{2}k^{2})+A_{1}\sigma_{0}\tau_{y}k_{z}+A_{2}({\bm \sigma}\times {\bm k})_{\hat{z}}\tau_{x}\;.
\end{equation}
Here the Pauli matrices $\tau$ ($\sigma$) act in the parity (spin) space, with $\tau_{0}$ and $\sigma_{0}$ the identity matrices, and $M$, $B_{i}$ and $A_{i}$ material-dependent parameters.
Since the topology of $H_{{\rm TI}}$
is determined by the sign of the ratio $M/B_{1}$, for simplicity and without loss of generality we assume that  $H_{{\rm t,b}}$ only differ from $H_{{\rm TI}}$ by the sign and magnitude of the mass term $M$. Hence $H_{t,b}$ are given by the same Eq.~\eqref{TIH} with the replacement  $M\rightarrow -m_{\mu}$, where $\mu=\pm$ labels  the top (bottom) insulator.

To implement the tunneling method we  rewrite the Hamiltonian, Eq.~\eqref{exacthamiltonian}, as a combination of the Hamiltonians for the top and the bottom interfaces plus an additional ``perturbative'' part connecting the two. As shown in Fig.~\ref{fig1}(a) this decomposition is
 \begin{equation}\label{decompositiion}
 H=H_{+}+H_{-}-H_{{\rm TI}}\,,
 \end{equation}
where the interface Hamiltonians are
\begin{subequations}\label{interfacesH}
  \begin{eqnarray}
&& H_+=H_{{\rm TI} }\Theta(a-z)+H_{{\rm t}}\Theta(z-a)\, \qquad\mbox{(top)}\,,
\label{semiinfinitesystms1}
\\
&&H_-=H_{{\rm TI}}\Theta(z+a)+H_{{\rm b}}\Theta(-a-z)\, \quad\mbox{(bottom)}\,.
\label{semiinfinitesystms2}
\end{eqnarray}
\end{subequations}Since translational invariance in the $x-y$ plane is preserved, the in-plane momentum ${\bm k}$ is a good quantum number. We solve the single interface problem for each ${\bm k}$ in Sec.~\ref{interfacestatessec} below.  Due to the change in topology across each interface,
at a given ${\bm k}$ there are four low-energy metallic states described by $H_\pm$, two for each interface. As the helicity operator, $\hat{h}=(\sigma\times {\bm k})\cdot\hat{z}/k$, commutes with the Hamiltonian~\cite{interfacesymmery,isaev} and has eigenvalues $\pm 1$, we can choose the interface states to have definite helicity, $\kappa=\pm$, so that
\begin{eqnarray}
H_{\mu}|\psi_{\mu,\kappa}\rangle &=& E_{\mu,\kappa}|\psi_{\mu,\kappa}\rangle\;.
\label{interface}
\end{eqnarray}
Here $E_{\mu,\kappa}$ are the energies of the helical states at the top and bottom interfaces.

We look for the eigenstates of the thin film as a linear combination of the top and bottom interface states,
\begin{equation}\label{wavefunctunn}
|\psi \rangle=\sum_{{\kappa,\mu=\pm}}{\alpha_{\mu,\kappa}|\psi_{\mu,\kappa}\rangle}\;,
\end{equation}
where $\alpha_{\mu,\kappa}$ are the coefficients to be determined. Substituting Eq.~\eqref{wavefunctunn} into the
Schr\"odinger equation, $H|\psi \rangle=E|\psi \rangle$,
using Eq.~\eqref{decompositiion}, and acting with $\langle\psi_{\mu',\kappa'}|$ on the left, we obtain a set of four linear equations for the
coefficients $\alpha_{\mu,\kappa}$:
    \begin{eqnarray}\label{t9}
    \nonumber
\sum_{{\kappa,\mu=\pm}} \Bigl\{&& \Delta E_{\mu,\kappa}\langle\psi_{\mu',\kappa'}|\psi_{\mu,\kappa}\rangle
 \\
 &&+\langle\psi_{\mu',\kappa'}| \Delta H_{\mu}|\psi_{\mu,\kappa}\rangle \Bigr\}\alpha_{ \mu,\kappa}=0\;.
\label{tb_gen}
    \end{eqnarray}
Here, $\Delta E_{\mu,\kappa}=E_{\mu,\kappa}-E$, $\Delta H_{\mu}=H_{\bar\mu}-H_{{\rm TI}}$, and $\bar\mu=-\mu$.
Non-trivial solutions exist when the determinant of the resulting matrix vanishes, thus giving the energy eigenvalues $E$, and the corresponding eigenvectors yield the wave functions in the basis of the single interface states. Further simplifications appear because of the piecewise constant parameter of the Hamiltonian in our problem.
Using $H_{\mu}$ from Eqs.~\eqref{interfacesH}, we find
\begin{subequations}\label{DeltaH_gen}
  \begin{eqnarray}
    &&\Delta H_+=H_{{\rm -}}-H_{{\rm TI}}=(H_{{\rm b}}-H_{{\rm TI}})\Theta(-a-z)\,,
    \\
    &&\Delta H_-=H_{{\rm +}}-H_{{\rm TI}}=(H_{{\rm t}}-H_{{\rm TI}})\Theta(z-a)\,.
  \end{eqnarray}
\end{subequations}The tight-binding nature of the method becomes explicit here since, as shown in Fig.~\ref{fig1}(b), for the film thickness comparable to or greater than the decay length of the interface states, in $\Delta H_\mu | \psi_{\mu,\kappa}\rangle$ the operator acts only on the decaying tails of the wave function, ensuring the smallness of the corresponding matrix element.

If we also recall that our model for the insulator Hamiltonian differs from the TI only by the sign of the mass, then Eq.~\eqref{DeltaH_gen} simplifies further, to
\begin{equation}
  \Delta H_\mu=-\sigma_{0}\tau_{z}(m_{\bar\mu}+M)\Theta(\bar\mu z-a)\,.
  \label{DeltaH_Mass}
\end{equation}

It is now evident that the problem of the free standing film requires careful consideration of the boundary conditions. Indeed, vacuum can be modeled~\cite{minfinity} by setting $m_\mu\rightarrow\infty$. In that case the wave functions vanish at surfaces, i.e. $\langle \bm r|\psi_{\mu, \kappa}\rangle=0$ for $\mu z-a>0$. Consequently, at first sight, there are no off-diagonal matrix elements, $\mu^\prime\neq \mu$, in the second term of Eq.~\eqref{tb_gen},
 $\langle\psi_{\mu',\kappa'}| \Delta H_{\mu}|\psi_{\mu,\kappa}\rangle$, simply because either the wave function or the operator vanishes everywhere in space. This result is clearly non-physical. However, since in this limit formally  $\Delta H_\mu \rightarrow -\infty$, we nominally have an infinitely large operator acting on the vanishing wave function. Hence it is obvious that the problem requires regularization in the large $m_\mu$ limit. Therefore below we evaluate these matrix elements for $m_\mu\gg M$, and show that they remain finite and independent of the value of $m_\mu$ as the vacuum limit is approached.

Using this method, we solve the thin film problem at each value of the in-plane momentum, ${\bm k}$, independently. The topological properties of the system are determined by the evolution in the structure of the eigenstates from $\bm k =0$ to large $\bm k$~\cite{fradkin}. It is therefore essential to accurately take into account the momentum dependence of both the single interface eigenstates, $|\psi_{\mu,\kappa}\rangle$, and the hybridization matrix elements in Eq.~\eqref{tb_gen}. Solving Eq.~\eqref{interface} with $B_2=0$, and then including this term perturbatively, as in Refs.~\onlinecite{thingap3,thingap2}, provides an adequate description of the states near the zone center, and an accurate evaluation of the gap, but leads to some erroneous conclusions about topological transitions.  Therefore below we derive the interface states keeping the full momentum dependence of the bulk Hamiltonian.

\subsection{Interface States}\label{interfacestatessec}

 To solve the single interface problem we follow the general approach of Ref.~\onlinecite{Rev2} as implemented in Ref.~\onlinecite{interfacesymmery}. We first find the exponential (along the $z$-axis) solutions of the bulk Hamiltonian, Eq.~\eqref{TIH}, select the eigenfunctions that decay away from the interface on each side, and then match the wave functions and their derivatives at the boundary.  As discussed above, we choose the wave functions to simultaneously be eigenstates of the helicity operator~\cite{isaev}. Using the label $\zeta=({\rm{I, TI}})$ for the I and TI sides respectively, the states at the top interface are given by~\cite{interfacesymmery}
\begin{equation}\label{statessTISE}
\psi_{\zeta,\kappa}(x,y,z)=\left(
                                   \begin{array}{c}
                                     ia_{\zeta,\kappa}(k) \\
                                     ib_{\zeta,\kappa}(k) \\
                                     \kappa a_{\zeta,\kappa}(k)e^{i\theta_{k}} \\
                                    \kappa b_{\zeta,\kappa}(k)e^{i\theta_{k}}  \\
                                   \end{array}\right)e^{i{\bm k}\cdot{\bm r}}e^{\lambda_{\zeta}(z-a)}\,.
\end{equation}
Here  $\kappa$ is the helicity eigenvalue,  the in-plane azimuthal angle $\theta_{k}=\tan^{-1}(k_{y}/k_{x})$, and
\begin{subequations}\label{asandbs}
  \begin{eqnarray}
    a_{\zeta,\kappa}(k) &=& A_{1}\lambda_{\zeta}-\kappa A_{2}k\;, \\
   b_{\zeta}(k) &=&  M_{\zeta}+B_{1}\lambda^{2}_{\zeta}-B_{2}k^{2}-E\;.
  \end{eqnarray}
\end{subequations}
The decay lengths $\lambda_{\zeta}$ satisfy the biquadratic equation
 \begin{equation}\label{lambda}
E^2-\mathcal{M}_{\zeta +}\mathcal{M}_{\zeta-}-A^{2}_{2}k^{2}=0\;,
\end{equation}
with $\mathcal{M}_{\zeta\pm}=M_{\zeta}+B_{1}\lambda^{2}_{\zeta}-B_{2}k^{2}\pm A_{1}\lambda_{\zeta}$, and $M_{{\rm TI}}=M$, $M_{{\rm I}}=-m$, see Fig.~\ref{fig1}. Since Eq.~\eqref{lambda} is biquadratic, there are two pairs of roots with positive  or negative real part. Therefore requiring the wave function to decay away from the interface (be normalizable) selects two allowed values for $\lambda_\zeta$, labeled by the index $\nu=\pm$, on each side of the interface. This index is then inherited by all the terms in Eq.~\eqref{statessTISE}, i.e. $a_{\zeta,\kappa}(k)\rightarrow a_{\zeta,\kappa,\nu}(k)$,  $b_{\zeta}(k)\rightarrow b_{\zeta,\kappa,\nu}(k)$ for each eigenstate $\psi_{\zeta,\kappa,\nu}$. The wave function of the interface state is given, at each side of the interface, by a linear combination of four eigenstates with different values of $\kappa$ and $\nu$.

In the absence of symmetry-breaking interface potentials~\cite{interfacesymmery}, helicity conservation allows us to solve for each value of $\kappa$ independently, i.e. on each side (I, TI) we look for solutions of the form
 \begin{equation}\label{lineacombstates}
        \Psi_{\zeta,\kappa}(x,y,z)= \sum_{\nu}{C_{\zeta,\kappa,\nu}\psi_{\zeta,\kappa,\nu}(x,y,z)}\;.
 \end{equation}
Here, $C_{\zeta,\kappa,\nu}$ are constants that are determined from the continuity of the wave function and its derivative at the boundary,  $\Psi_{{\rm TI}, \kappa }(x,y,a)=\Psi_{{\rm I},\kappa}(x,y,a)$ and  $\partial_{z}\Psi_{ {\rm TI}, \kappa }(x,y,a)=\partial_{z}\Psi_{{\rm I},\kappa}(x,y,a)$. For the top interface we find
$E_{+,\kappa}=\kappa A_{2}k$, and
\begin{subequations}\label{lambdasa2k}
\begin{eqnarray}
 \lambda_{{\rm TI},\nu}(k) &\equiv&\lambda_{\nu}(k) = \frac{A_{2}+\nu\sqrt{A^{2}_{1}-4B_{1}M_{k}}}{2B_{1}}\;,
 \label{lambda_TI} \\
  \lambda_{{\rm I},\nu}(k)  &\equiv& -\Lambda_{\nu}(k) = -\frac{\nu A_{2}+\sqrt{A^{2}_{1}+4B_{1}m_{k}}}{2B_{1}}\;,
\end{eqnarray}
\end{subequations}
with $M_{k}=M-B_{2}k^2$ and $m_{k}=m+B_{2}k^2$. Substituting $E_{+,\kappa}$ and $\lambda_{\zeta,\nu}$ in Eq.~\eqref{asandbs} gives the spinor structure of the interface state, and
requires $C_{{\rm I},\kappa,+}=0$ to match the spinors at $z=a$. The remaining three coefficients satisfy
\begin{subequations}\label{boundarycoefs}
\begin{eqnarray}
 C_{{\rm I},\kappa,-}&=&C_{{\rm TI },\kappa,+}\frac{\lambda_{-}(k)-\lambda_{+}(k) }{ \lambda_{-}(k)+\Lambda_{-}(k)}\;,  \\
C_{{\rm TI },\kappa,-}&=&-C_{{\rm TI },\kappa,+}\frac{\lambda_{+}(k)+\Lambda_{-}(k)}{\lambda_{-}(k)+\Lambda_{-}(k)}\;,
\end{eqnarray}
\end{subequations}
as well as the normalization condition.

Since we are interested here in the surface states of a TI we take the limit of a large gap insulator, $m\rightarrow\infty$,  so that $\Lambda_{-}\approx \sqrt{m/B_{1}}\gg |\lambda_{\pm}|$, and expand the coefficients in $\lambda_\pm/\Lambda_-$ to find
\begin{subequations}\label{approximatecoefs}
\begin{eqnarray}
   C_{{\rm TI },\kappa,+}&\approx& C_{0}(k)\left(1-\frac{\lambda_{+}(k)}{\Lambda_{-}}\right)\equiv c_{+}\;,\\
   C_{{\rm TI },\kappa,+}&\approx&-C_{0}(k)\left(1-\frac{\lambda_{-}(k)}{\Lambda_{-}}\right)\equiv c_{-}\;,  \\
     C_{{\rm I},\kappa,-}&\approx& -C_{0}(k)\frac{\lambda_{+}(k)-\lambda_{-}(k)}{\Lambda_{-}}\equiv c_{{\rm I}}\;,\\
     C_{0}(k)&=&\sqrt{\frac{A_{1}M_{k}}{2(A^{2}_{1}-4B_{1}M_{k})}}\;.
\end{eqnarray}
\end{subequations}
With this, the wave functions of the top interface in Fig.~\ref{fig1}, that are eigenfunctions of Eq.~\eqref{semiinfinitesystms1}, are given by
\begin{equation}\label{topsattessetise}
\Psi_{+, \kappa} =f_{+, \kappa}( k,z-a)\left(
                       \begin{array}{c}
                         i \\
                         i \\
                         \kappa e^{i\theta_{k}} \\
                        \kappa e^{i\theta_{k}} \\
                       \end{array}
                     \right)e^{i{\bm k}\cdot{\bm r}}\;,
\end{equation}
where
\begin{eqnarray}
  f_{+, \kappa}( k,z)&=&\Bigl( \Bigl[c_{+}e^{\lambda_{+}(k)z}+c_{-}e^{\lambda_{-}(k)z}\Bigr]\Theta(-z)
  \nonumber
  \\
  &&\qquad +c_{{\rm I}}e^{-\Lambda_{-}(k)z}\Theta(z)\Bigr)\,.
\end{eqnarray}

To find the topological eigenfunctions of the bottom interface, Hamiltonian Eq.\eqref{semiinfinitesystms2}, we follow an identical procedure, and find  $E_{-,\kappa}=-\kappa A_{2}k$,  and
\begin{equation}\label{statesbottim}
\Psi_{-, \kappa} =f_{-, \kappa}(k,z+a)\left(
                       \begin{array}{c}
                         i \\
                         -i \\
                         \kappa e^{i\theta_{k}} \\
                         -\kappa e^{i\theta_{k}} \\
                       \end{array}
                     \right)e^{i{\bm k}\cdot{\bm r}}\;,
\end{equation}
with
\begin{eqnarray}
  f_{-, \kappa}(k,z) =&&\Bigl( \Bigr[c_{+}e^{-\lambda_{+}(k)z}+ c_{-}e^{-\lambda_{-}(k)z}\Bigr]\Theta(z)
  \nonumber
  \\
  &&+c_{I}e^{\Lambda_{-}(k)z}\Theta(-z)\Bigr)\,.
\end{eqnarray}Eq.~\eqref{topsattessetise} and Eq.~\eqref{statesbottim} give the eigenfunctions for the top and bottom interface in a I-TI-I junction in the limit of a large gap insulator. As expected, these states are linearly dispersing, until they merge with the continuum at $k_{s}=\sqrt{M/B_{2}}$. They also are purely helical, with states of equal energy having opposite helicity at the top and bottom interfaces. Since the spinors of the opposite helicity states are orthogonal, this restricts the possible matrix elements in the tunneling Hamiltonian for the film, Eq.~\eqref{tb_gen}. The decay lengths ($\lambda^{-1}_{\pm}(k)$) of the interface states depend on the in-plane momentum, which means that the tunneling matrix elements also depend on  ${\bm k}$, opening the possibility for a non-trivial band structure of the electronic states in the film. We now proceed to determine these bands.

\subsection{Effective Hamiltonian for Thin Film }\label{effectivehamiltoniansec}

We use the solutions for single interfaces, Eqs.~\eqref{topsattessetise} and~\eqref{statesbottim}, to determine the matrix elements in Eq.~\eqref{tb_gen}. The spinor structure of the interface states has no coordinate dependence, and therefore the convolution of the spinors and the integration of the wave functions can be performed separately. It is easy to check that the states with different helicities at the same surface, as well as any states at opposite surfaces, are orthogonal, so that
\begin{equation}\label{diagonal}
  \langle\psi_{\mu^\prime,\kappa^\prime}|\psi_{\mu,\kappa}\rangle= \delta_{\mu^\prime\mu}\delta_{\kappa^\prime\kappa} \;.
\end{equation}
The remaining elements of the tunneling Hamiltonian are
\begin{equation}\label{t24}
  t_{\mu^\prime\kappa^\prime\mu\kappa}(k,a)\equiv\langle\psi_{\mu',\kappa'}| \Delta H_{\mu}|\psi_{\mu,\kappa}\rangle\,.
\end{equation}
The convolution of the spinors in Eq.~\eqref{topsattessetise} and Eq.~\eqref{statesbottim} with the $\sigma_0\tau_z$ matrix in $\Delta H_\mu$ from Eq.~\eqref{DeltaH_Mass} restricts these elements to be off-diagonal, connecting the states at opposite surfaces, with identical helicities, so that only $t_{\mu,\kappa,-\mu,\kappa}\ne 0$. Setting $m_+=m_-=m\gg M$ for the free standing film as above, we find
\begin{equation}\label{offdiagonal1}
t_{\mu,\mu',\kappa,\kappa'}(k,a)=\delta_{\mu^\prime,-\mu}\delta_{\kappa^\prime\kappa}t(k,a)\;,
\end{equation}
where $t(k,a)$ is the overlap (hopping) integral obtained by integrating the functions $f_{\pm,\kappa}(k,z)$ from Eqs.~\eqref{topsattessetise} and \eqref{statesbottim} with the Heaviside function in the definition of $\Delta H_\mu$ in Eq.~\eqref{DeltaH_Mass}. The result is
\begin{equation}
t(k,a)=\frac{4m}{\Lambda^{2}_{-}}C^{2}_{0}(k)\left(e^{-2a\lambda_{-}}-e^{-2a\lambda_{+}}\right)(\lambda_{-}-\lambda_{+})\;.
\end{equation}
The necessity of keeping the gap (mass term) of the topologically trivial insulator finite now becomes clear: for a vacuum termination, $m\gg M$, we find $m/\Lambda^{2}_{-}=B_{1}$, independent of $m$. Quite generally the same result would hold for a real interface with a large gap insulator. From the definition of $\lambda_{\pm}(k)$ in Eq.~\eqref{lambda_TI} the explicit functional form of $t(k, a)$ depends on whether $\varsigma(k)=A^{2}_{1}-4B_{1}M_{k}$ is positive or negative, and yields
\begin{equation}\label{thopping}
t(k,a)= \frac{4A_1 M_k e^{-a \frac{A_1}{B_1}}}{\sqrt{|\varsigma(k)|}}\times
        \begin{cases}
            \sin\left(\frac{\sqrt{|\varsigma(k)|}a}{B_{1}}\right) & \text{if $\varsigma(k) < 0$}\;, \\
            \sinh\left(\frac{\sqrt{|\varsigma(k)|}a}{B_{1}}\right) & \text{if $\varsigma(k)>0$ }\;. \\
        \end{cases}
\end{equation}
As expected, the hopping element is exponentially dependent on the thickness of the film. However, if $\varsigma(k) < 0$, it also exhibits oscillatory behavior as a function of both thickness and the in-plane momentum.

These oscillations are critical for the determination of the topological nature of the bands in the subsequent section, so we address them briefly here. In the $\varsigma(k)<0$ regime, the decay lengths $\lambda_\pm$ are complex, so that the wave function of the interface state oscillates as well as decays away from the boundary. The hopping integral, $t(k,a)$, inherits this behavior. Recall that  $M_{k}=M-B_{2}k^{2}$, and that for topological insulators $MB_1>0$. In that case the existence of the oscillations depends on the sign of $\varsigma(0)= A^{2}_{1}-4B_{1}M$, which is negative for the well investigated and proposed TI materials Bi$_{2}$Se$_{3}$, Bi$_{2}$Te$_{3}$ and Sb$_{2}$Te$_{3}$ according to the parameters of the ${\bm k}\cdot{\bm p}$  analysis of Ref.~\onlinecite{TIhamiltonian}. For these compounds the tunneling matrix element does depend non-monotonically on the thickness for the momenta near the $\Gamma$-point of the surface Brillouin Zone. Recall also  that the surface states merge with the bulk bands at $k_s=\sqrt{M/B_2}$, when $M_k=0$, and therefore $\varsigma(k) > 0$. It follows that for these materials there is another characteristic momentum, $k_{0}=\sqrt{(4B_{1}M-A^{2}_{1})/(4B_{1}B_{2})}< k_s$, such that $t(k>k_0, a)$ is always positive and non-oscillatory. This conclusion could only be reached because we did not carry out a perturbative expansion in $k$, but kept $B_2k^2$ term in the solution of the interface problem.

We are now in the position to write the Hamiltonian for the thin film in the (rearranged) spinor basis of the surface states, $(\widehat\psi_{+,+}\ , \widehat\psi_{+,-}\ , \widehat\psi_{-,-}\ , \widehat\psi_{-,+})$, where  (again see Eqs.~\eqref{topsattessetise} and ~\eqref{statesbottim}) $\widehat{\psi}_{\mu,\kappa}=\frac{1}{2}(i,\mu i,\kappa e^{i\theta_k},\kappa\mu e^{i\theta_k})^{(T)}$, and $(T)$ means transposed.
Introducing the Pauli matrices acting in the helicity ($\kappa$) space, $\Gamma_{i}$, we find
\begin{equation}\label{effectivH1}
H_{{\rm tn}}=\left(
               \begin{array}{cc}
                A_{2}k\Gamma_{z} & t(k,a)\Gamma_x \\
                t(k,a)\Gamma_x & A_{2}k\Gamma_{z} \\
               \end{array}
             \right)\;.
\end{equation}
The eigenvalues give the doubly degenerate (index $\eta$ below) bands in the TI thin film,
\begin{equation}\label{disperssion}
E_{\pm,\eta}=\pm\sqrt{(A_{2}k)^{2}+t(k,a)^2}\;.
\end{equation}

\subsection{Energy dispersion for films of candidate materials}\label{spectrumsec}

\begin{figure}[t]
    \begin{center}
            \subfigure{
            \includegraphics[width=0.99\columnwidth]{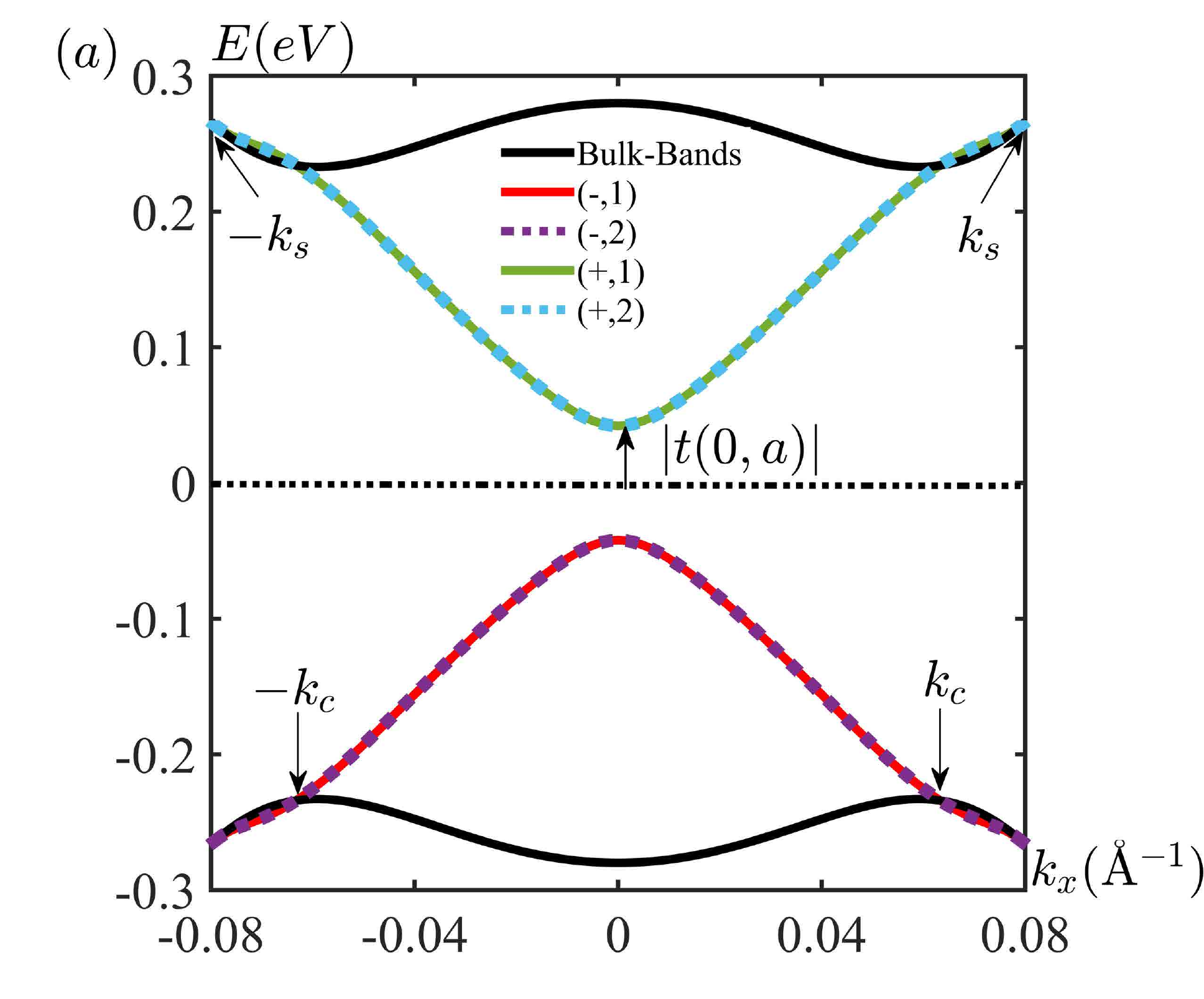}}
        \subfigure{
        \includegraphics[width=0.88\columnwidth]{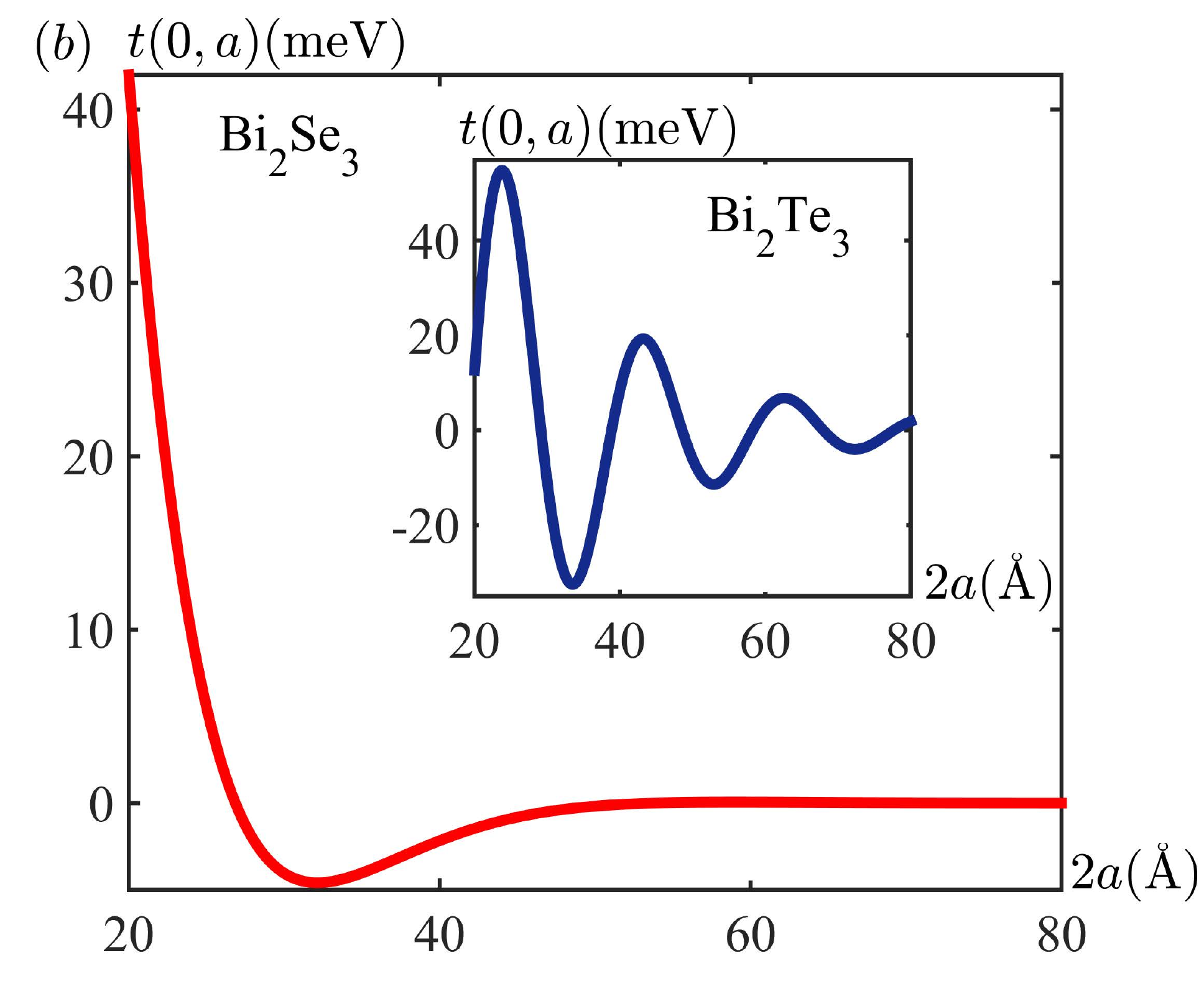}}
    \end{center}
    \caption{Band structure of the topological insulator thin film. (a) Dispersion relation of a TI thin film of thickness $2a=2$ nm, made of Bi$_{2}$Se$_{3}$, with ${\bm k}\cdot{\bm p}$ parameters $A_{1}=2.26\; eV$\AA , $A_{2}=3.33\; eV$\AA,
$B_{1}=6.86\; eV$\AA$^{2}$, $B_{2}=44.5\; eV$\AA$^{2}$, and
$M=0.28\ eV$ from Ref.~\onlinecite{TIhamiltonian}.
The thin film bands merge with the bulk bands of the TI at $k_{c}$, while in the absence of hybridization the surface states merge with the bulk at $k_{s}$. (b) Thickness dependence of the gap $t(0,a)$ for Bi$_{2}$Se$_{3}$.
Inset: Thickness dependence of the gap $t(0,a)$ for Bi$_{2}$Te$_{3}$, with ${\bm k}\cdot{\bm p}$ parameters $A_{1}=0.3\; eV$\AA , $A_{2}=2.87\; eV$\AA,
$B_{1}=2.79\; eV$\AA$^{2}$, $B_{2}=57.38\; eV$\AA$^{2}$, and
$M=0.3\ eV$ from Ref.~\onlinecite{TIhamiltonian}.}
\label{fig2}
\end{figure}

 The characteristic energy dispersion Eq.~\eqref{disperssion}, obtained for the parameters of Bi$_2$Se$_3$, is shown in Fig.~\ref{fig2}(a). The spectrum is gapped, with the direct gap at ${\bm k}=0$ given by
\begin{equation}\label{t01}
t(0,a)=\frac{4A_{1}Me^{-a\frac{A_{1}}{B_{1}}}\sin\left(\frac{\sqrt{|A^{2}_{1}-4B_{1}M|}a}{B_{1}}\right)}{\sqrt{|A^{2}_{1}-4B_{1}M|}}\;.
\end{equation}
For the commonly  known TIs, Bi$_{2}$Se$_{3}$, Bi$_{2}$Te$_{3}$ and Sb$_{2}$Te$_{3}$, the ratio $A_1/B_1\sim 0.1 - 1$\AA$^{-1}$. Hence for realistic thicknesses of a few quintuple layers, $a A_1/B_1>1$, and  the exponential factor significantly suppresses this induced gap compared to the bulk gap value, generally to values on the order of 10meV or less, as shown.

In Fig.~\ref{fig2}(a) it is important to note that value of the in-plane momentum, $k$,  at which the TI thin film bands merge with the bulk energy bands is different from that for a single interface. The bulk band energy at $k_z=0$ is $E_{B}=\sqrt{(A_{2}k)^{2}+M^{2}_{k}}$ and therefore the merging points are given by the solutions to $t(k,a)=\pm M_{k}$. One of the roots of this equation is always $k_{s}=\sqrt{M/B_{2}}$ found for the surface state. However, using Eq.~\eqref{thopping}, we find another root, $k_{c}$, from the solution of the equation
$\sin q= c q$ or $\sinh q=c q$,
where $q=\sqrt{|\varsigma(k)|}a/B_{1}$, the parameter-dependent constant $c=(B_1/4a A_1)\exp(a A_1/B_1)$, and the choice of the function ($\sin$ or $\sinh$) depends on the sign of $\varsigma(k)$ as discussed above. The equation has the solution for $\sin q$ if $c<1$, and for $\sinh q$ if $c>1$. From the estimate above, the exponential factor is large, and we are in the latter regime. It follows that $k_c>k_0$, always in the range where there are no oscillations of the tunneling matrix element. We also find that generically $k_c<k_s$, providing the natural cutoff wave vector for our low-energy Hamiltonian.

\begin{figure}[t]
    \begin{center}
            \subfigure{
            \includegraphics[width=0.48\columnwidth]{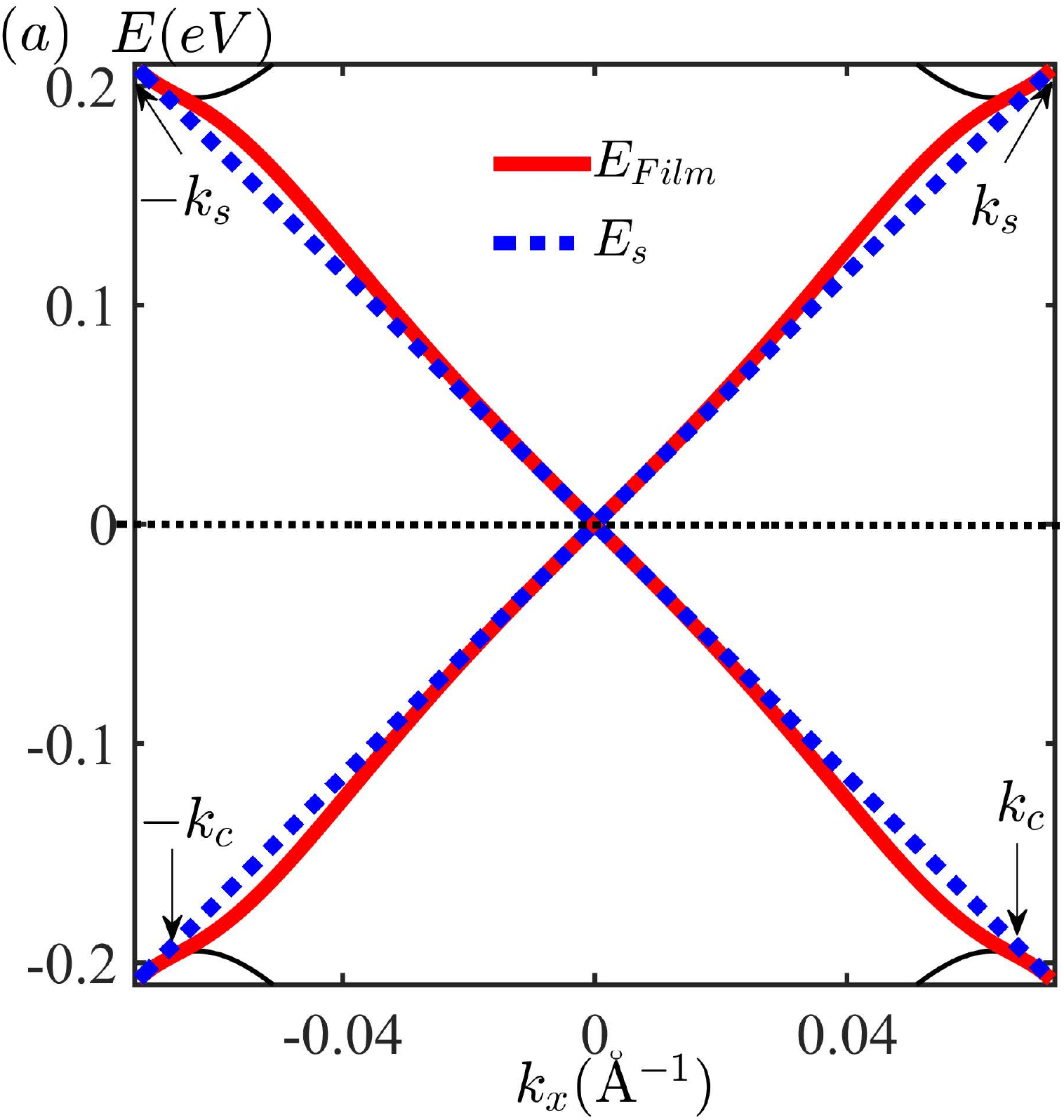}}
        \subfigure{
        \includegraphics[width=0.48\columnwidth]{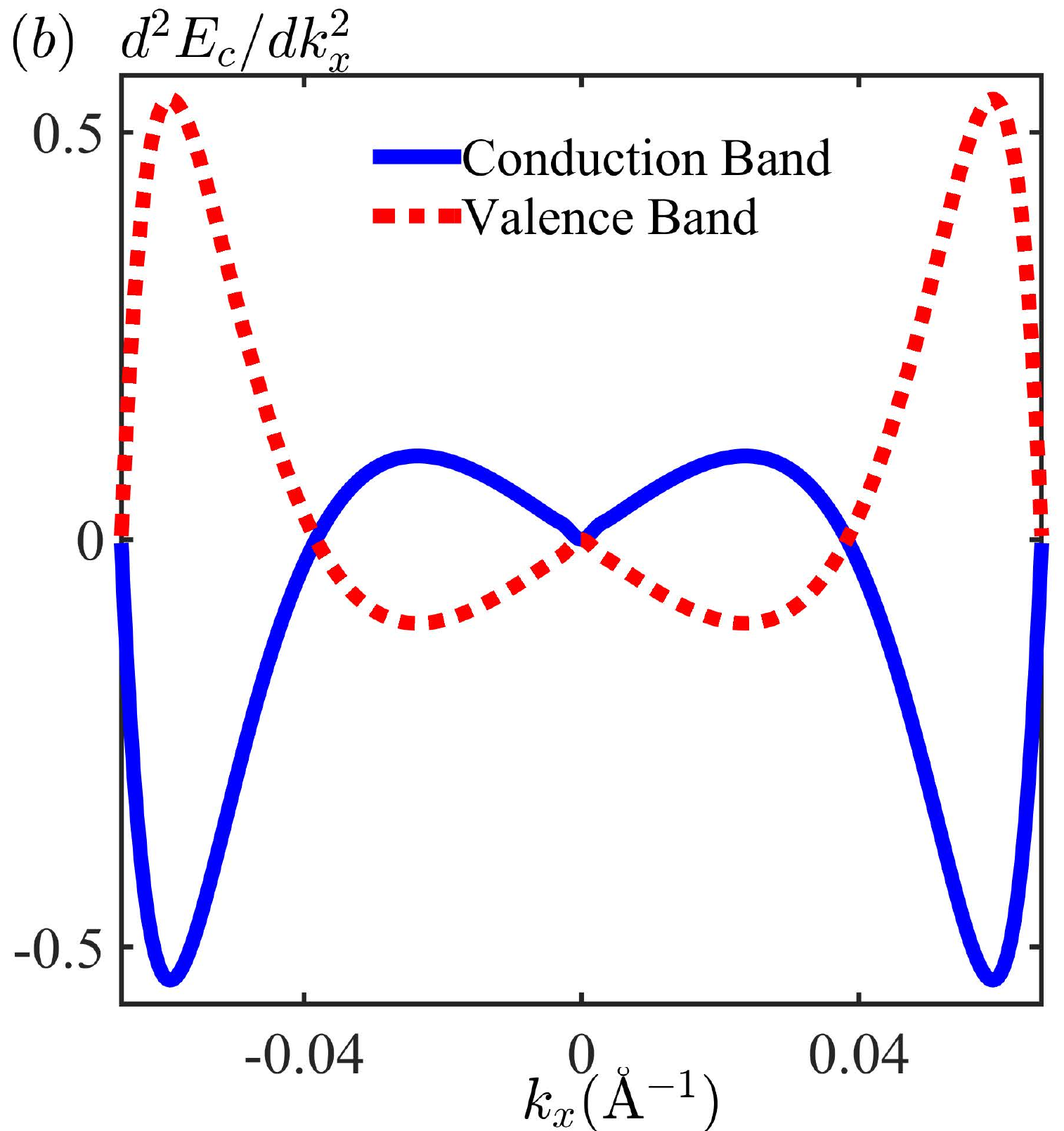}}
    \end{center}
    \caption{Gapless states in the thin film of a topological insulator. (a) Dispersion relation for Bi$_2$Te$_3$ for thickness $2a=29.14 $ \AA\;such that $t(0,a)=0$. Note the deviations from the linear dispersion of the decoupled surface states (dashed line). (b) The band curvature as a measure of the deviation from the linear dispersion (which would exhibit no curvature) of the surface state.}
\label{newfig3}
\end{figure}

The oscillations of the energy gap as a function of film thickness are shown in Fig.~\ref{fig2}(b). According to the ${\bm k}\cdot{\bm p}$ parameters, the oscillations are much more pronounced in Bi$_2$Te$_3$, which would make it a suitable candidate for the experimental observation of the thickness dependence of the gap.

There are film thicknesses for which $t(0,a)$ vanishes, and the states are gapless and remain approximately semi-metallic near ${\bm k}=0$. However, it would be erroneous to interpret the gapless spectrum as an indication that the states at both surfaces are decoupled. As is clear from Eq.~\eqref{thopping}, the hopping integral $t(k, a)$ does not vanish at all $k$ simultaneously, so that the surfaces remain coupled even in this case. This is illustrated in Fig.~\ref{newfig3}, where we show that the energy dispersion for the film remains non-linear even when the gap closes, and plot the band curvature as a measure of the deviation from the linearly dispersing surface state.

This observation illustrates a crucial point for our subsequent discussion. The gap magnitude, $|t(0, a)|$, that we obtain is in quantitative agreement with the findings of Refs.~\onlinecite{thingap1,thingap2,thingap3}. However, the curvature of the bands at larger ${\bm k}$ is controlled by the momentum dependence of the tunneling parameter $t(k,a)$ and thus differs very substantially from the perturbative, in ${\bm k}$, results in Refs.~\onlinecite{thingap2,thingap3}. The band curvature plays a critical role in the analysis of the possible topological phases in this system, an issue that we turn to in the next section in which we investigate the topological properties of the electronic states in a TI thin film.

\section{Topological Phases in a TI thin Film}

Both the gap in the energy spectrum of the TI thin film,  and the band curvature of the low-energy states are controlled by the hybridization function $t(k, a)$. As this function may exhibit sign changes as a function of its arguments, it is appropriate to ask whether topological electronic phases exist in this system. We answer this question using two different methods.

\subsection{Pseudospin textures}\label{massparamiter}

  We first analyze the topological phases of the thin film by exploring their connection to the pseudo-spin winding in momentum space. Since the bands of the Hamiltonian are always doubly degenerate, we have the freedom to choose a convenient basis by combining the wave functions of the degenerate states. We make our choice based on two requirements.

  First, we demand that the basis states should be the eigenstates of the Hamiltonian at ${\bm k}=0$. From Eq.~\eqref{effectivH1} these are the symmetric and antisymmetric combinations of the states with the same helicity at the top and bottom interfaces.
  Second,
  we choose the basis so that in the limit of a thick film (in which the surfaces decouple, with $t\rightarrow 0$) the Hamiltonian regains the well-known helical Dirac form, $(\bm\sigma\times {\bm k})_z$. This amounts to returning from the helicity to the spin space according to $\psi_{\mu, \uparrow}=i\left(\widehat\psi_{\mu,+}\ +\widehat\psi_{\mu,-}\right)$, and $\psi_{\mu, \downarrow}=e^{i\theta_k}\left(\widehat\psi_{\mu,+}\ -\widehat\psi_{\mu,-}\right)$, where we remind the reader that $\mu=\pm$ indicates the upper and the lower surfaces of the film. This step introduces the even and odd combinations of the states with the opposite helicities at the same interface. Together, they suggest using the basis
 $\widetilde\psi=(\psi_{ T,\uparrow}+\psi_{B,\uparrow}\ , -\psi_{T,\downarrow}+\psi_{B,\downarrow}\ , -\psi_{T, \uparrow}+\psi_{B, \uparrow}\ ,  \psi_{T, \downarrow}+\psi_{B,\downarrow})^{(T)}$, in which the Hamiltonian takes the block-diagonal form
\begin{equation}\label{hamiltonian11}
\mathcal{H}({\bm k})=\left(
                \begin{array}{cc}
                  \mathcal{H}_{1}({\bm k}) & 0 \\
                  0 & \mathcal{H}_{2}({\bm k}) \\
                \end{array}
              \right)\;,
\end{equation}
with
\begin{subequations}\label{h1h2}
\begin{eqnarray}
 \mathcal{H}_{1}({\bm k}) &=& \left(
              \begin{array}{cc}
                t(k,a) & -A_{2}(k_{y}+ik_{x}) \\
                -A_{2}(k_{y}-ik_{x}) & -t(k,a) \\
              \end{array}
            \right)\;,
   \\
 \mathcal{H}_{2}({\bm k}) &=& \left(
              \begin{array}{cc}
                -t(k,a) & -A_{2}(k_{y}+ik_{x}) \\
                -A_{2}(k_{y}-ik_{x}) & t(k,a) \\
              \end{array}
            \right)\;.
\end{eqnarray}
\end{subequations}
 Note that within each block the upper and lower components of the basis functions have a definite and opposite $z$-component of the spin $(\uparrow,\downarrow)$, but describe different (odd vs even) combinations of the two interface states. Consequently, if we introduce the Pauli matrices $\Sigma_i$  acting within each block, they do not correspond to physical spin operators. Defining  $\alpha_i$ as the Pauli matrices in the band (blocks of ${\cal H}$) space, spin is given by $S_z=\hbar\alpha_0\Sigma_z/2$, and $S_{x,y}=\hbar\alpha_x\Sigma_{x,y}/2$, confirming the pseudo-spin nature of the vector $\bm \Sigma$. Nonetheless, this pseudo-spin can be used for the topological classification of the states.

Each block ${\cal H}_{1,2}$ is identical to the well-studied hamiltonian of a Chern insulator~\cite{bernevig}, with the tunneling matrix element $t(k,a)$ playing the role of a mass, so that the bulk spectrum consists of two bands separated by an energy gap $2|t(0, a)|$. Additionally, if the sign of the mass term changes between ${\bm k}=0$ and large values of ${\bm k}$,  each block supports a gapless edge mode. The direction of the propagation of this edge mode is determined by the sign of the mass, and hence the two blocks yield two counterpropagating states. If the basis functions for the two blocks were spin eigenstates, the edge would carry a net spin current, and Eq.~\eqref{hamiltonian11} would describe a spin Hall insulator. Since they are not, the TI thin film generally supports a pseudospin-$\bm \Sigma$ Hall effect, which is reminiscent of the valley Hall effect in graphene~\cite{valleyHallgrpahen}. We discuss in Sec.~\ref{edgestatesec} the special circumstances under which this corresponds to the physical spin-Hall effect.

\begin{figure*}[t]
    \begin{center}
            \subfigure{
            \includegraphics[width=40mm,height=35mm ]{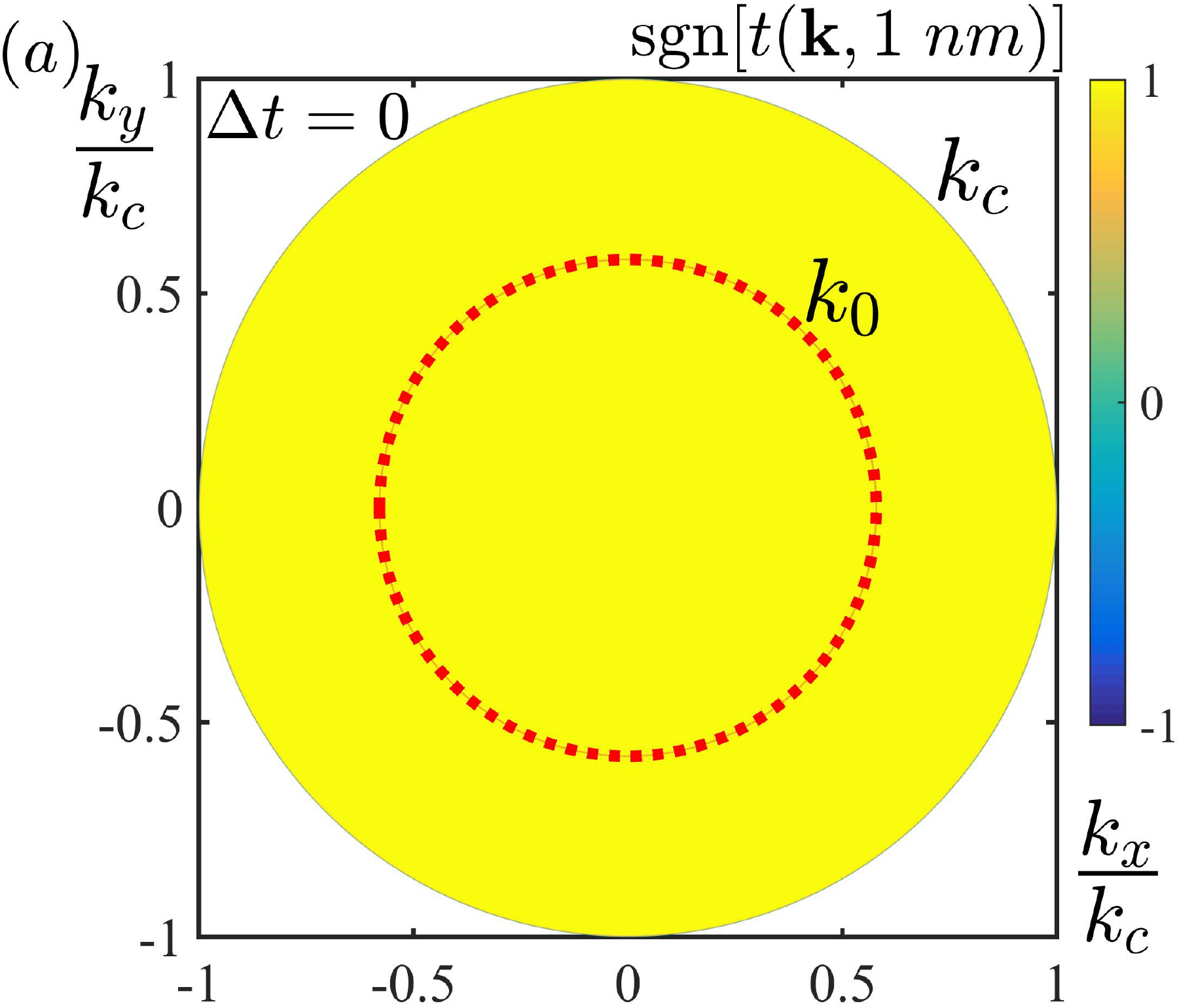}}%
        \subfigure{
            \includegraphics[width=40mm,  height=35mm ]{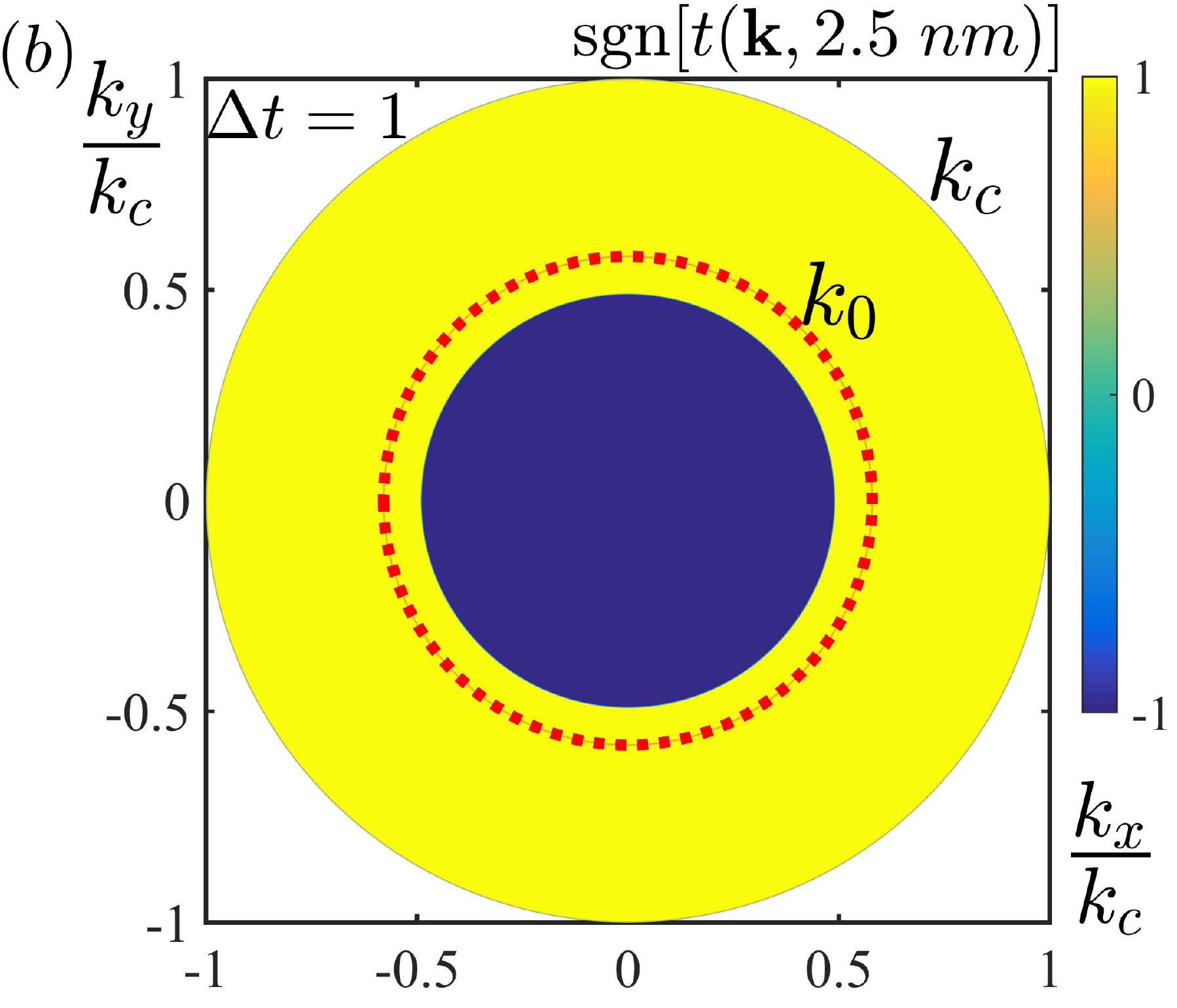}}%
                               \subfigure{
           \includegraphics[width=40mm, height=35mm]{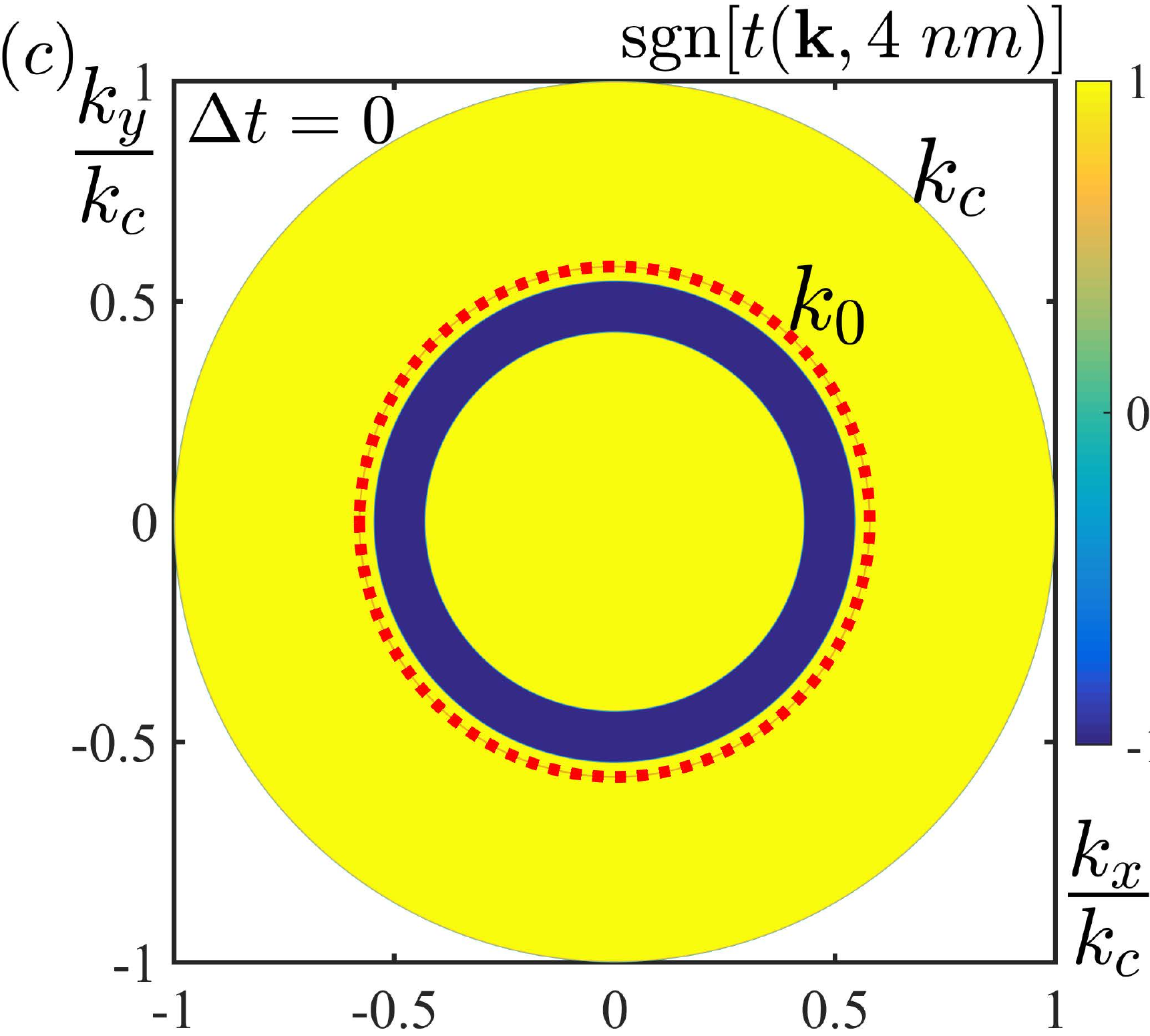}}%
                   \subfigure{
          \includegraphics[width=40mm, height=35mm]{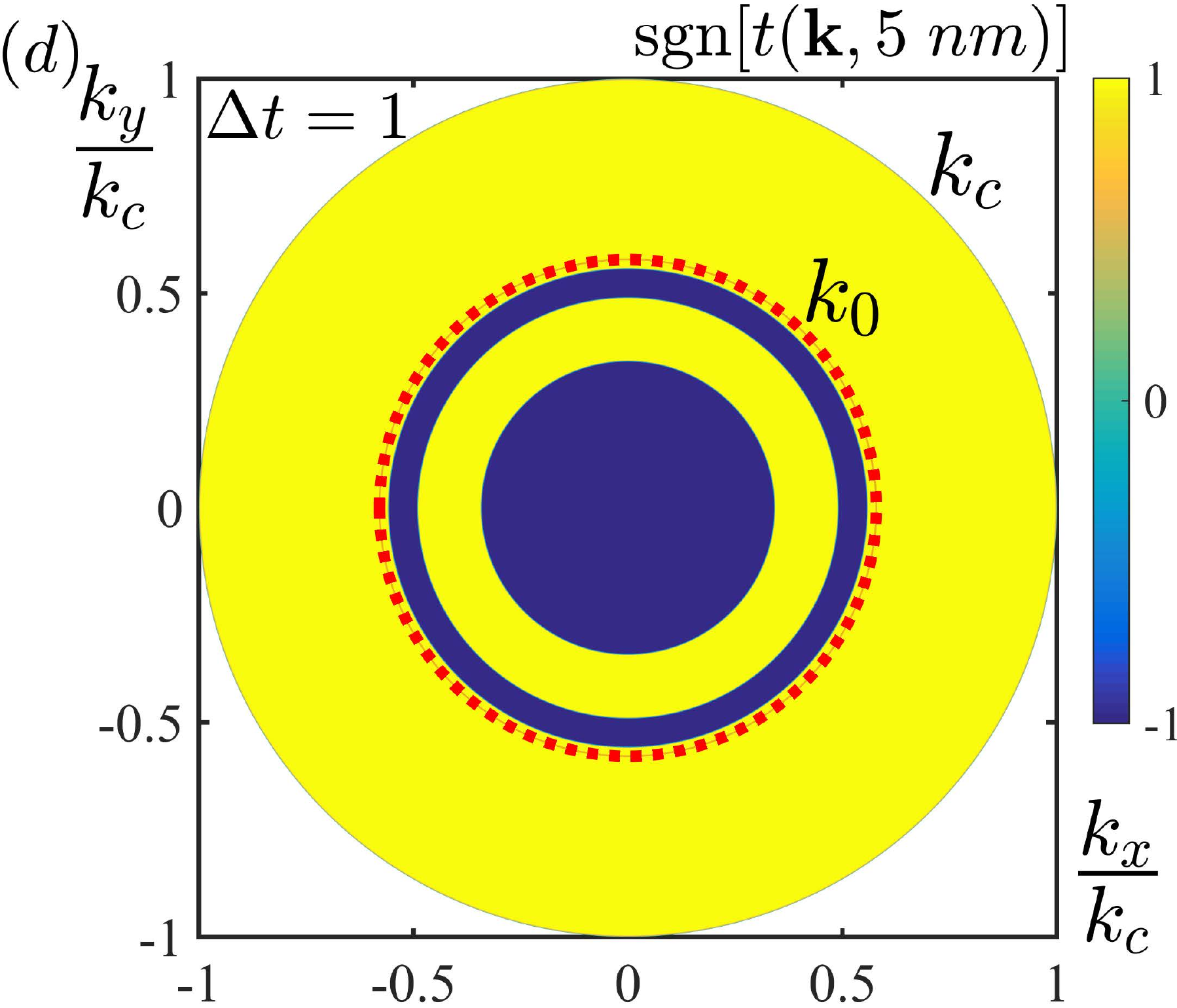}}
                      \subfigure{
            \includegraphics[width=40mm,height=35mm ]{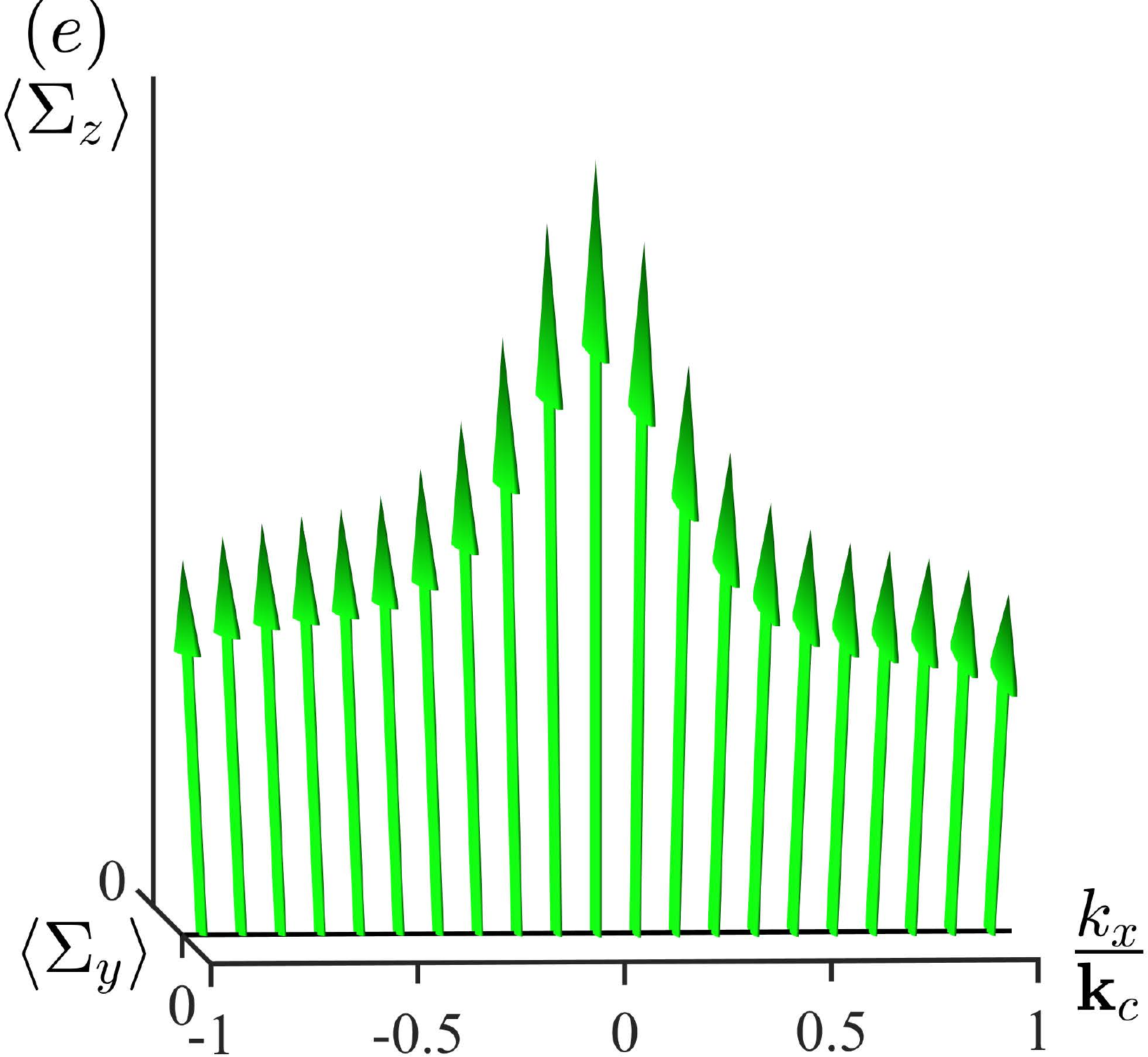}}%
        \subfigure{
            \includegraphics[width=40mm,  height=35mm ]{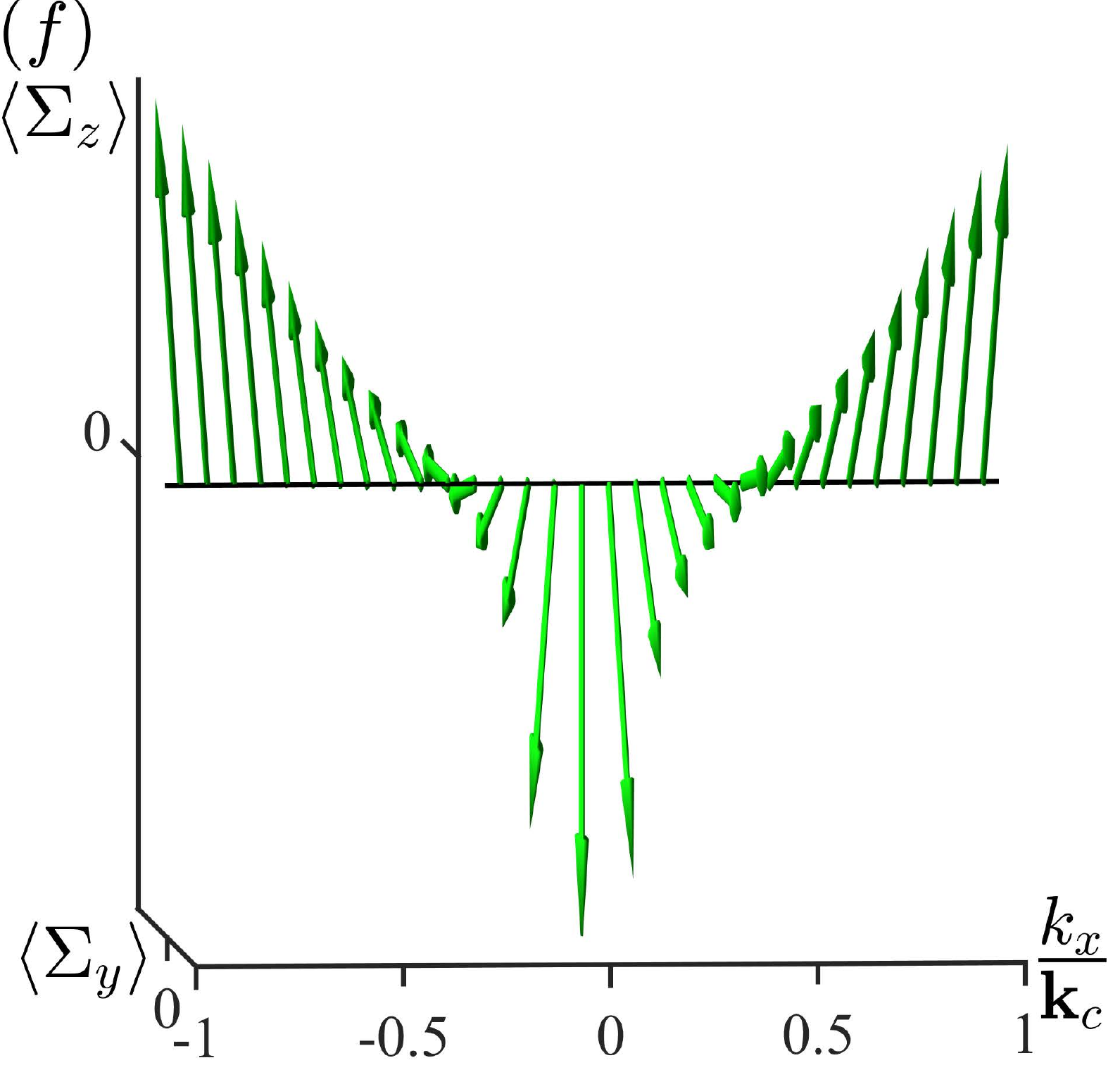}}%
                               \subfigure{
           \includegraphics[width=40mm, height=35mm]{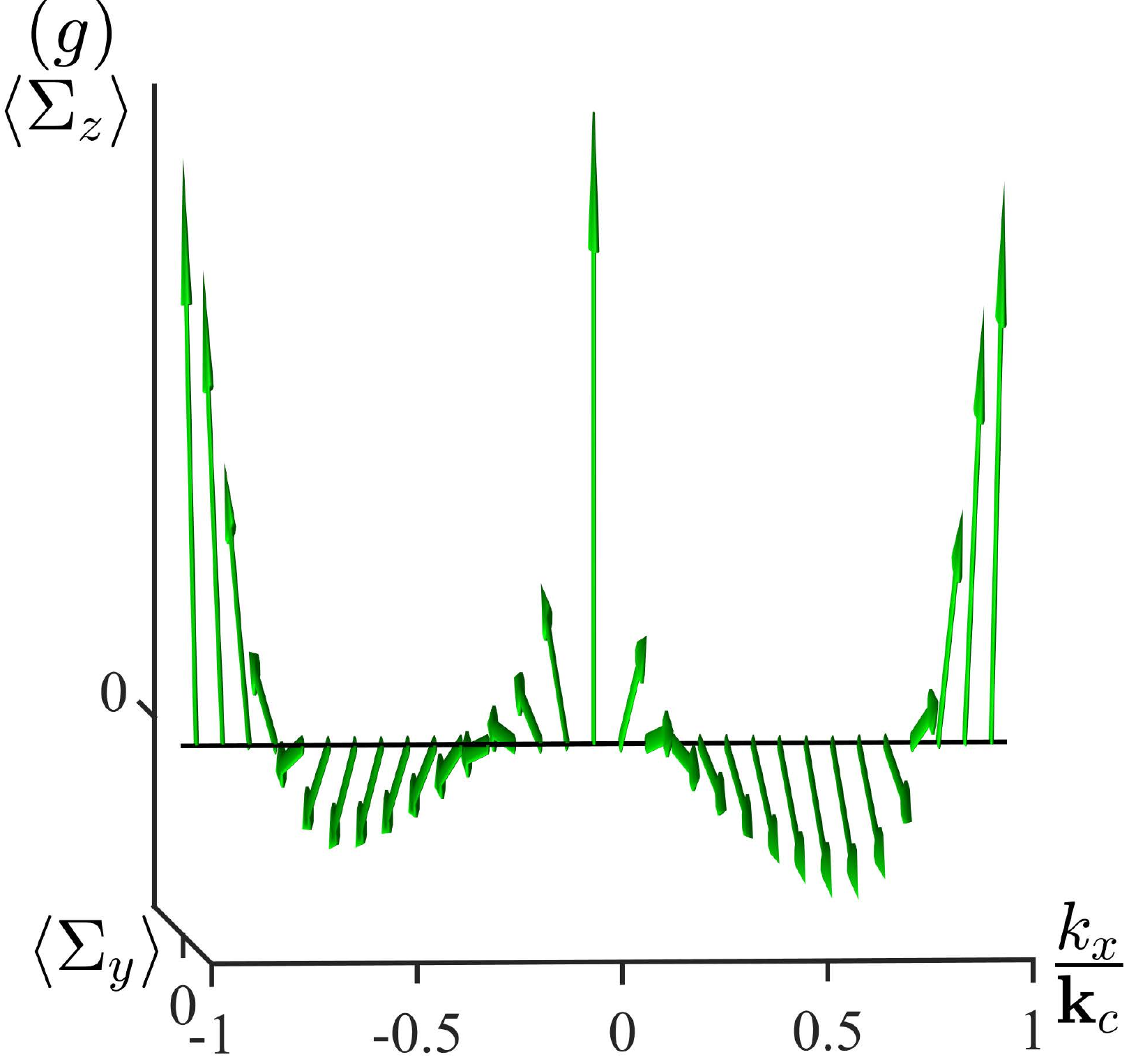}}%
                   \subfigure{
          \includegraphics[width=40mm, height=35mm]{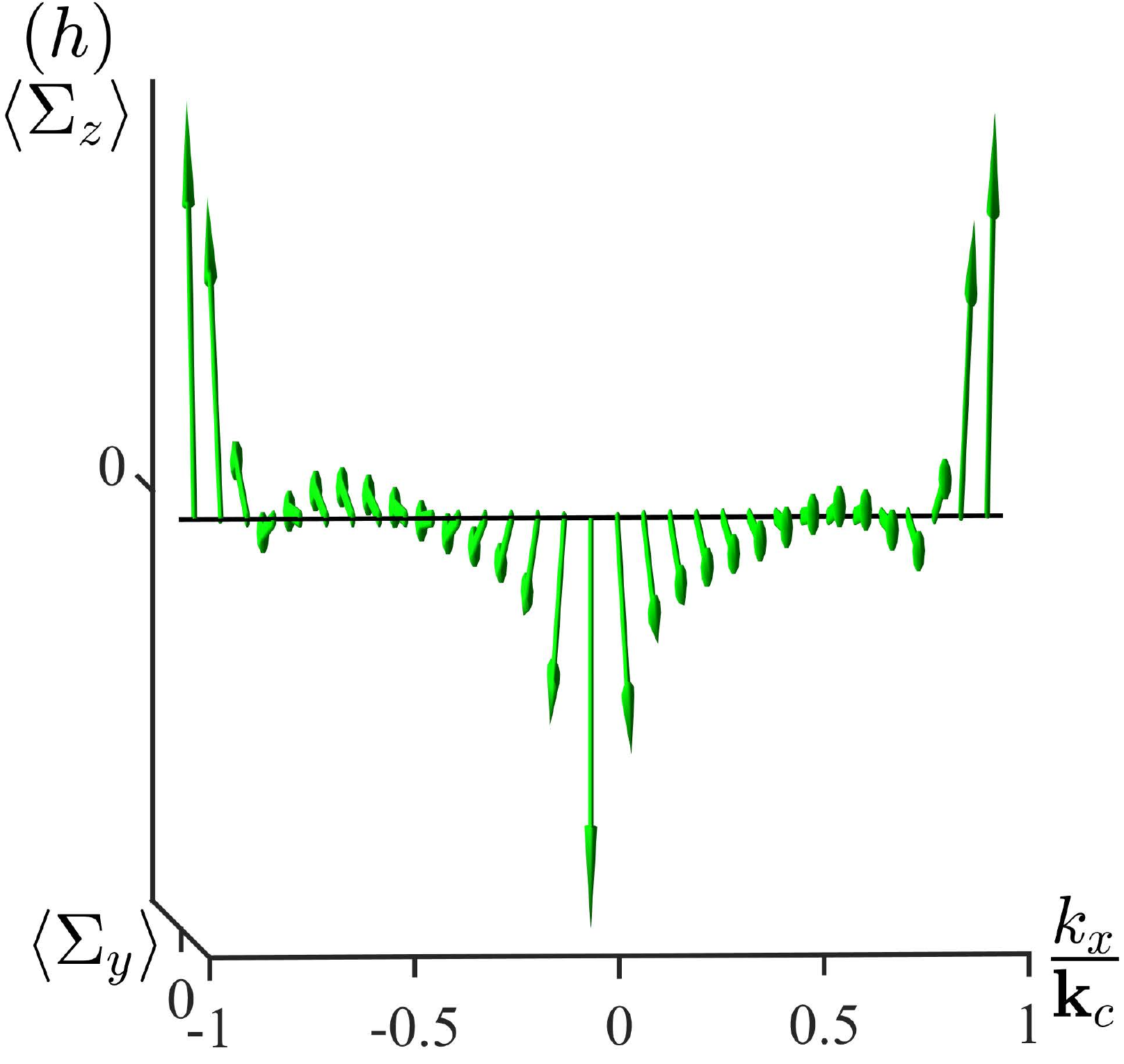}}
    \end{center}
    \caption{The top row shows the sign of the mass term, $t(k,a)$, as a function of momentum up to the cutoff $k_c$ for the band parameters of Bi$_{2}$Se$_{3}$, listed in the caption to Fig.~\ref{fig2}. The four panels show different film thickness, $2a$.
    (a) In a $2$ nm thick film there is no sign change, so that the index defined in Eq.~\eqref{signchange} $\Delta t=0$, and the system is topologically trivial; (b) For a thickness of $5$ nm, the mass term displays a single sign change, hence $\Delta t=1$, and the structure is topologically non-trivial; (c) An $8$ nm film has two sign changes, so that $\Delta t=0$. (c) For $a=10$ nm, there are  three sign changes, so $\Delta t=1$. (e)-(h) pseudo-spin textures of the valence band states, $E_{-,1}$, along the line $k_y=0$ for the cases (a)-(d). Only panels (f) and (h), corresponding to $\Delta t=1$  in panels (b) and (d) respectively, have a non trivial winding characteristic of topological states. }
\label{fig:fig2}
\end{figure*}

 The number of sign changes of the mass term, $t(k, a)$, depends on the film thickness and the sign of $\varsigma(0)=A_1^2-4B_1 M$, see Eq.~\eqref{thopping}. For $\varsigma(0)>0$ the system is always topologically trivial. In the opposite case, $\varsigma(0) < 0$ (realized for Bi$_{2}$Se$_{3}$, Sb$_{2}$Te$_{3}$ and Bi$_{2}$Te$_{3}$), the sign of $t(k=0, a)$, and the number of sign changes of the mass term between ${\bm k} =0$ and the oscillation cutoff $k=k_0$ defined in Sec.~\ref{effectivehamiltoniansec} depends on the film thickness. Illustrative cases are shown in Figs.~\ref{fig:fig2}(a)-(d).
If the number of sign changes of $t(k, a)$ is odd, the system is topological. Hence the sign of the tunneling matrix element at $\bm k=0$ defines the ``topological mass index'' of the film,
\begin{equation}\label{signchange}
\Delta t=\frac{[1-{\rm sgn}(t(0,a)]}{2}\,,
\end{equation}
which  depends on the parameters of the TI material and its thickness. If $\Delta t=1$ ($\Delta t=0$), the TI thin film is in the topological (non-topological) regime.

To confirm that the index $\Delta t$ defines the topological properties of the thin film we consider the winding of the pseudo-spin, ${\bm \Sigma}$, in momentum space~\cite{bernevig,TKNN}. Since our system has time-reversal and inversion symmetry, it is sufficient to consider the pseudo-spin winding in only one of the bands, and we choose here the bottom (valence) band of ${\cal H}_1$, labeled as ``$-,1$'' hereafter. The pseudo-spin content of the states in other bands (``$+, 1$'';``$-,2$'', and ``$+,2$''  in obvious notations) is related to the band we consider via $\langle\Sigma_{i}\rangle_{+,1}=-\langle\Sigma_{i}\rangle_{-,1}$, and $\langle\Sigma_{i}\rangle_{\pm,1}=\langle\Sigma_{i}\rangle_{\mp,2}$.

The components of the pseudo-spin of the eigenstates of the film, $\langle\Sigma_{i}\rangle$, as a function of $k_x$, at $k_{y}=0$, are shown in Figs.~\ref{fig:fig2} (e)-(h). Here we use the the same set of parameters as in Figs.~\ref{fig:fig2} (a)-(d).  For that cut in momentum space $\langle\Sigma_{x}\rangle_{-,1}=0$, so that we only plot  $\langle\Sigma_{z}\rangle_{-,1}$  and $\langle\Sigma_{y}\rangle_{-,1}$. Our system is rotationally invariant, so that the entire pseudo-spin texture can be inferred from these panels. We find a non-trivial winding, with a skyrmion-like texture, whenever $t(0,a)<0$, confirming our analysis.

A comparison of Figs.~\ref{fig:fig2} (b) and (d) shows that, even though $t(k,a)$ goes through a different number of sign changes in these two examples of topological states, the net pseudo-spin winding angle is the same for both cases. This is not {\it a priori} obvious, and naively one could expect an extra $2\pi$ phase for each additional ``domain wall'' of the mass term in momentum space. This observation is relevant for the determination of the edge states that are a signature of the topological phases. The main question that arises is whether all such phases of the thin film have a single edge state per band, or whether it is possible for the cases with multiple sign changes to generically have an odd number of edge states (even though the topological protection may extend only to a single state that remains gapless after hybridization). To answer this question we directly compute the Chern number that gives the number of edge modes.

\subsection{ Chern Number of TI Thin Film}\label{chernumbersec}

\begin{figure}[h]
  \centering
  \includegraphics[scale=0.45]{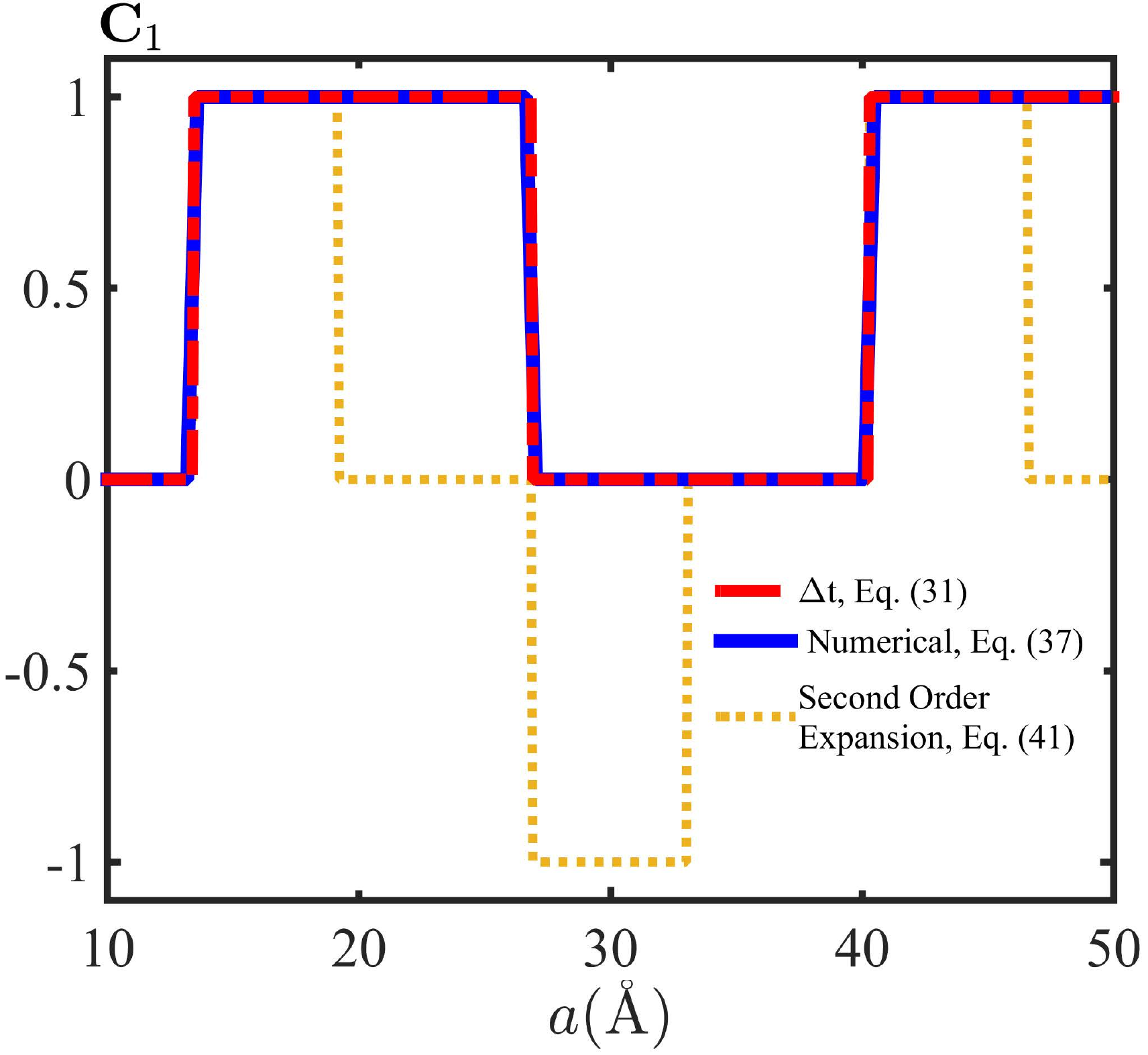}
  \caption{ Chern number of the filled valence band from the upper block of the Hamiltonian, Eq.~\eqref{hamiltonian11}, $\mathbf{C}_{1}$, as a function of the film thickness. $\mathbf{C}_{1}=0$ and $|\mathbf{C}_{1}|=1$ indicate the trivial and the topological insulating states respectively. Note the agreement between the numerical result and the index based on the sign of the tunneling mass term at zero momentum.
  The second order expansion of $t(k, a)$ gives spurious topological transitions with no closing of the gap, see text. }\label{fig3}
\end{figure}

Our next task is to compute the Chern number of a TI thin film. As discussed above, each block of the Hamiltonian,  Eq.~\eqref{hamiltonian11}, describes a two band insulator, with the two blocks related by time-reversal symmetry, which results in doubly-degenerate valence and conduction bands. Therefore the Chern number for each block is the topological index that characterizes its phases~\cite{bernevig}.

We first rewrite each $2\times 2$ Hamiltonian in Eq.~\eqref{h1h2} via the pseudo-spins $\Sigma$ coupled to a pseudo-Zeeman field ${\bm h}({\bm k})$:
\begin{equation}\label{htwolevel}
\mathcal{H}_{i}({\bm k})=h_{x}({\bm k})\Sigma_{x}+h_{y}({\bm k})\Sigma_{y}+(-1)^{i+1} h_{z}({\bm k})\Sigma_{z}\;,
\end{equation}
here $i=1,2$ is the block index, and the vector ${\bm h}({\bm k})=(-A_{2}k_{y},A_{2}k_{x},t(k,a))$. The Berry curvature of each block is \cite{fradkin}
\begin{equation}\label{curvatute2band}
\Omega_{xy,i}({\bm k})=(-1)^{i+1}\frac{1}{2}\epsilon_{abc}\frac{\partial \hat{h}_{a}({\bm k})}{\partial k_{x}}\frac{\partial \hat{h}_{b}({\bm k})}{\partial k_{y}}\hat{h}_{c}({\bm k})\;,
\end{equation}
where the unit vector $\widehat{{\bm h}}({\bm k})={\bm h}({\bm k})/|{\bm h}({\bm k})|$,  and $\epsilon_{abc}$ is the Levi Civita tensor. Therefore $\Omega_{xy,1}({\bm k})=-\Omega_{xy,2}({\bm k})$, and the net Berry curvature of the entire system ($\Omega_{xy,1}({\bm k})+\Omega_{xy,2}({\bm k})$) vanishes identically for all ${\bm k}$, as required by TR and inversion symmetry~\cite{niuberryphase}.

Although the net Berry curvature vanishes, the Berry curvature for each block does not always vanish. The corresponding block Chern number is
\begin{equation}\label{chern2}
\mathbf{C}_{i}=\frac{1}{2\pi}\int_{0}^{{\bm k}_{c}}d^{2}{\bm k}\; \Omega_{xy,i}({\bm k})\,,
\end{equation}
where $k_{c}$ is the momentum cutoff. For a non vanishing Chern number, $\mathbf{C}_1=-\mathbf{C}_2=j\neq 0$, the system hosts $j$ pairs of time reversed versions of the quantum Hall state, associated with counter circulating TRI edge states. It is therefore sufficient to compute the Chern number of just one block, and hereafter we focus on the upper, $i=1$, block of Eq.~\eqref{hamiltonian11}.
It is also important to note from Eq.~\eqref{chern2} that the TI thin film system is not a spin quantum Hall insulator even in the topological phase (in contrast to the claims in Refs.~\onlinecite{thingap3,thingap2}), since the counter circulating edge states carry opposite band index, $i$, but not opposite spin. We will discuss this issue more extensively below.

Due to the non-monotonic dependence of $t(k,a)$ on momentum it is convenient to calculate the Berry curvature and the Chern number using the numerically-efficient method described in Ref.~\onlinecite{numericalchern}. In this method momentum is discretized, ${\bm k}_{l}=(k_{x,l_{x}},k_{y,l_{y}})$, where $k_{x,l_{x}}=k_{c}l_{x}/N_{x}$,  $k_{y,l_{y}}=k_{c}l_{y}/N_{y}$,  the indices $l_{x}=(0,\ldots, N_{x})$, $l_{y}=(0,\ldots, N_{y})$, and  $N_{x}$ and $N_{y}$ define the step of the grid.  For a gapped system the Berry connection is defined via the link variables of the occupied  (valence band) states,
 \begin{equation}\label{linkvariable}
U_{i}=\frac{\langle\psi({\bm k}_{l})|\psi({\bm k}_{l}+\hat{{\bm e}}_{i})\rangle}{|\langle\psi({\bm k}_{l})|\psi({\bm k}_{l}+\hat{{\bm e}}_{i})\rangle|}\,,
 \end{equation}
where $\hat{{\bm e}}_{x}$ and $\hat{{\bm e}}_{y}$ are the unit vectors along the $x$ and $y$ directions. Within this formalism the discrete distribution of the Berry curvature takes the form
\begin{equation}\label{berrycurvnum}
\Omega_{xy,1}({\bm k}_{l})=\ln\left(\frac{U_{x}({\bm k}_{l})U_{y}({\bm k}_{l}+\hat{{\bm e}}_{x})}{U_{x}({\bm k}_{l}+\hat{{\bm e}}_{y})U_{y}({\bm k}_{l})}\right)\,,
\end{equation}
and the Chern number is given by
\begin{equation}\label{chernnumnum}
\mathbf{C}_{1}=\frac{1}{(2i\pi)}\sum_{l}{\Omega_{xy,1}({\bm k}_{l})}\,.
\end{equation}

We plot $\mathbf{C}_{1}$ as a function of Bi$_2$Se$_3$ film thickness in Fig.~\ref{fig3}, and show that its behavior is identical to that of the topological mass index, $\Delta t$.
In agreement with our analysis of the phase winding of the pseudo-spin vector, the Chern number in the topological phase is 1, indicating that we have only one edge state per band, i.e. two counter-circulating edge modes in total, irrespective of the number of sign changes of the mass parameter $t(k, a)$. The phase boundaries between topologically distinct phases, corresponding to $t(0,a)=0$, are always gapless, with linearly dispersing Dirac-like states at low energies. Note however that the top and bottom surface states remain coupled as is evident from the band curvature, (see Fig.~\ref{newfig3} and Sec.~\ref{spectrumsec}).

We now review the approximation used in Refs.~\onlinecite{thingap3,thingap2}, which amounts to expanding $t(k,a)$, to second order in $k$,
\begin{equation}\label{tsecondorder}
t(k,a)\approx t(0,a)-B(a)k^2\,,
\end{equation}
where
\begin{eqnarray}
\label{Bofa}
B(a)&=&\frac{4A_{1}B_{2}e^{-a\frac{A_{1}}{B_{1}}}F(a)}{|\varsigma(0)|^{3/2}}\,,
\\
\label{fb}
F(a)&=&2aM\sqrt{|\varsigma(0)|}\cos\left(\mathcal{X}(a)\right)-\varsigma(0)\sin\left(\mathcal{X}(a)\right)\;.
 \end{eqnarray}
Here $\mathcal{X}(a)=\sqrt{|\varsigma(0)|}a/B_{1}$ and, as before, $\varsigma(0)=A^{2}_{1}-4B_{1}M$. Note that $B(a)$ is an oscillating function. Within this approximation setting the cutoff ${\bm k}_{c}\rightarrow\infty$ in Eq.~\eqref{chern2} we find for the Chern number
\begin{equation}\label{secondorderchern}
\mathbf{C}_{1,2}=\mp\frac{1}{2}\left[{\rm sgn}(t(0,a))+{\rm sgn}(B(a))\right]\;,
\end{equation}
with the upper sign corresponding to $\mathbf{C}_{1}$.
In this approximation the topological transitions occur at the points where the sign of $t(0,a)$ changes as in previous discussion as well as at the thicknesses where $B(a)$ changes sign. This is  shown in Fig.~\ref{fig3}, is in complete accord with Ref.~\onlinecite{thingap2,thingap3}, but does not agree with the results of the calculation presented above. The main point of contrast between the two methods is the implication of the topological transition without a gap closing at $B(a)=0$, which we believe to be an artifact of using the expansion. In contrast, the topological phase diagram obtained from the exact form of $t(k,a)$ reflects the non-adiabatic connection between the two distinct topological phases as they are separated by a linearly dispersing Dirac semi-metallic phase~\cite{diracsemimetal1}.

\section{Edge states in TI thin Film}\label{edgestatesec}
Having shown that a TI thin film can exhibit topological phases, we now look for the gapless edge states as the signatures of the topological nature of the film.
We consider a semi-infinite thin film occupying the $y>0$ half-plane, and solve the eigenvalue equation for the state localized near the edge. Because of the block-diagonal nature of the Hamiltonian in Eq.~\eqref{hamiltonian11}, it is sufficient to solve for the wave functions satisfying each of the blocks separately.
Translational invariance along the surface makes $k_{x}$  a good quantum number, while we need to replace $k_{y}$ by the momentum operator $-i\partial_{y}$. Without loss of generality we focus on the upper block of the Hamiltonian, and solve the problem in two steps.

\subsection{Zero energy states}
\label{sec_egde_zero}
For trivial hard-edge boundary conditions, when $\psi(x,y=0)=0$, the Hamiltonian is particle-hole symmetric, and we expect that the edge state exists at $E=0$ for $k_x=0$. In this case
the eigenvalue problem becomes
  \begin{equation}\label{eignevlaue}
    \left(
      \begin{array}{cc}
       t(-i\partial_{y},a) & iA_{2}\partial_{y} \\
        iA_{2}\partial_{y} & -t(-i\partial_{y},a) \\
      \end{array}
    \right)
    \widetilde{\phi}_{1}(y)=0\;,
\end{equation}
where $\widetilde{\phi}_{1}(y)$ is a spinor in the pseudospin $\bm\Sigma$ space.  For the states localized near the edge we take an ansatz $\widetilde{\phi}_{1}(y)=\phi_{1}e^{\lambda y}$, which gives
 \begin{equation}\label{lambda1}
[  t(-i\lambda,a)\Sigma_{z} +i\lambda A_{2}\Sigma_{x}]\phi_{1}=0\,.
\end{equation}
Multiplying by $-i\Sigma_{x}$, we obtain the equation
 \begin{equation}\label{lambda2}
 t(-i\lambda,a)\Sigma_{y}\phi_{1} =\lambda A_{2}\phi_{1}\,,
\end{equation}
which requires $\phi_{1}$ to be proportional to an eigenstate of the $\Sigma_{y}$ matrix, $\Sigma_{y}\phi_{\pm}=\nu\phi_\pm$ with $\nu=\pm 1$.
In turn, $\lambda$ must satisfy
\begin{equation}\label{lambdaequation}
 [t(-i\lambda,a)]^2-(A_{2}\lambda)^{2}=0\,.
\end{equation}
Our strategy now is as follows. Assume we have a hard wall boundary condition, i.e. $\widetilde{\phi}_{1}(0)=0$. Then for the existence of the normalizable edge state we must have (at least) two roots, $\lambda_{1,2}$, of Eq.~\eqref{lambdaequation} with
${\rm Re}(\lambda_{1,2})<0$ corresponding to the same eigenvalue of $\Sigma_y$, i.e. both satisfying $t(-i\lambda,a)=\nu \lambda A_{2}$ with a given $\nu$. In that case the solution has the form
\begin{equation}\label{phi1}
\phi_{1}(y)=C_{1}\phi_{\nu}(\exp(\lambda_1 y)-\exp(\lambda_2 y))\,,
\end{equation}
where $C_1$ is a constant.

In general, the transcendental equation for $\lambda$, Eq.~\eqref{lambdaequation}, is not analytically solvable. In the regime where the second order expansion in $\bm k$ and the exact results agree,  we make the corresponding expansion in $\lambda$, so that $t(-i\lambda,a)=t(0,a)+B(a)\lambda^2$. In this approximation the solutions equation for $\lambda$ are
\begin{equation}\label{lambdaquadratic}
\lambda_{\nu,\pm}=
\pm\frac{\nu A_{2}
\pm\sqrt{A^2_{2}-4 B(a)t(0,a)}}{2B(a)}\,.
\end{equation}
For $B(a)t(0,a)>0$ the two values of $\lambda$ for the same $\nu$ have the same sign of the real part, and therefore can be combined to give the non-trivial edge state in Eq.~\eqref{phi1}. Therefore the requirement for the existence of the edge state is for $t(0,a)$ and $B(a)$ to have the same sign, in agreement with Eq.~\eqref{secondorderchern}, which only holds in the perturbative regime.

The decay constants for the bottom block of the Hamiltonian are obtained from the equation above by replacing $\nu\rightarrow -\nu$. Consequently, for our geometry, the general solution for the $E=0$ edge eigenstate in the regime $t(0,a)<0$, $B(a)<0$, where the expansion gives the correct topological phase, has the form
\begin{equation}\label{edgestaequad}
\phi(y)=\left(\begin{array}{c}
          C_{1}\phi_{+} \\
           C_{2}\phi_{-}
        \end{array}\right)\left[e^{\lambda_{+,-}y}-e^{\lambda_{+,+}y}\right] \;.
\end{equation}The result of such an expansion, while qualitatively correct, is somewhat misleading. This becomes obvious if we take advantage of
the exponential smallness of $t(0,a)$ and $B(a)$ in the film thickness, $t(0,a)B(a)\propto \exp(-2aA_{1}/B_{1})$, to expand the square root in Eq.~\eqref{lambdaquadratic} and find that our values are
\begin{equation}\label{smalllambdas}
 \lambda_{\nu,\pm}=\nu\left\{\frac{t(0,a)}{A_{2}}, \frac{A_{2}}{B(a)}\right\}\,.
\end{equation}
The first of these is a small number, but the second value is exponentially large, and falls outside regime of validity of the small ${\bm k}$ expansion, and, indeed, may fall beyond the cutoff for the low-energy theory, $k_c$.
We also numerically solved Eq.~\eqref{lambdaequation} and found two pairs of roots: one at $\pm t(0,a)/A_2$ within numerical accuracy, and the other, complex, with
$|\lambda_{c\pm}|>k_c$. Only in the topological regime, $t(0,a)<0$, do two roots with the same sign of the real part correspond to the same eigenvalue $\nu$, and therefore only in this regime we can have an edge state at zero energy. However, once again, one of the roots falls beyond the limit of applicability of the low energy theory developed here.

We are therefore in the situation where the topological indices and the approximate calculations unequivocally point towards the existence of the edge states, but the {\it exact} behavior of the wave functions of such states can only be determined from the theory that includes the high-energy physics close to the edges of the Brillouin zone. Our conclusion here is that the wave function is of the form given in Eq.~\eqref{phi1}, with the exponentially small $\lambda_1\approx t(0,a)/A_2$, and $\lambda_2\gg \lambda_1$. In the next section, we use this observation to develop a linearized theory of the edge states.

\subsection{Linearized approximation and the dispersion of the edge states}

The wave function of the edge state, Eq.~\eqref{phi1},  in the limit of $\lambda_2/\lambda_1\gg 1$ has a peak at $y_0\approx |\lambda_2|^{-1}\ln (\lambda_2/\lambda_1)$. That point separates the ``boundary layer'' that contains the fraction $\sim 2(\lambda_1/\lambda_2)\ln (\lambda_2/\lambda_1)\ll 1$ of the total weight of the edge state from the long-tail decay $\propto \exp(\lambda_1 y)$ where most of the quasiparticle weight is contained. We now follow Ref.~\onlinecite{isaev} and develop a linearized theory describing these tails.

Since $t({\bm k}, a)$ depends on the momentum only quadratically, to linear order in momentum the Hamiltonian, Eq.~\eqref{hamiltonian11}, is
 \begin{equation}\label{linear1}
\mathcal{H}({\bm k})=-A_{2}\alpha_{0}({\bm \Sigma}\times {\bm k})_{z}-|t(0,a)|\alpha_{z}\Sigma_{z}\,.
\end{equation}
Here $\alpha_i$ and $\Sigma_i$ are the Pauli matrices operating in the block and pseudospin space, respectively, as defined in Sec.~\ref{massparamiter}, and we emphasize that we work in the topological regime by explicitly writing $t(0, a)<0$. The key formal point is that, since this Hamiltonian is linear in the momentum, we can no longer impose the continuity of both the wave function and its derivative at the edge. The underlying physical reason for this, of course, is that we are only solving the problem on the long length scale $|\lambda_1|^{-1}$, beyond the thin boundary layer. Within this layer the wave function varies rapidly satisfying the exact boundary conditions of the full Hamiltonian. This complex behavior can be absorbed into the linearized low energy theory~\cite{isaev,falko,akhmerov,basko} by imposing a boundary condition that is consistent with particle conservation (so that the resulting Hamiltonian is self-adjoint in the half-space), time-reversal, and inversion symmetry. This boundary condition takes the form ${\cal B}\psi\vert_{\rm{edge}}=0$, where the matrix ${\cal B}= \alpha_{0}\Sigma_{0}-i{\cal M}\hat{\bm n}\cdot(\hat{z}\times{\bm \Sigma})$, and
${\cal M}$ commutes with time reversal, anticommutes with the current normal to the boundary, $\{\hat{\bm n}\cdot\alpha_{0}{(\hat{z}\times\bm\Sigma)},{\cal M}\}=0$ and is Hermitian and unitary, so that ${\cal M}={\cal M}^\dagger$, and ${\cal M}^{2}=1$. Here $\hat{\bm n}$ is the unit vector normal to the edge.

For our example of a film occupying the $y>0$ half-plane, the matrix ${\cal B}$ takes the form
\begin{equation}\label{bmatrix}
{\cal B}=\alpha_{0}\Sigma_{0}+\alpha_{z}\left[\Sigma_{z}\sin\vartheta+\Sigma_{y}\cos\vartheta\right]\;.
\end{equation}
Here the single parameter $\vartheta$ depends on the details of the boundary conditions, and controls the particle-hole symmetry breaking at the edge~\cite{isaev}. The particle-hole symmetric situation (as with the usual hard wall boundary) corresponds to $\vartheta=0$. However, in general the one-parameter family of boundary conditions, Eq.~\eqref{bmatrix}, describes all possible choices of ${\cal B}$ for which ${\cal H}$ is self-adjoint since
the von Neumann deficiency index of the problem is $n_{d}=1$.~\cite{AdjointsBook} Consequently, by varying $\vartheta$ we are able to describe a wide range of edge states corresponding to different boundary choices in the original Hamiltonian.

Since ${\cal B}$ is diagonal in the band index, we find solutions for each of the blocks of the Hamiltonian in the form
\begin{equation}\label{edgestateslinear}
\psi_{1}(x,y)=\left(
       \begin{array}{c}
        \mathcal{U}_{1}\\
         {\bm 0} \\
       \end{array}
     \right)e^{ik_{x}x}e^{\lambda_{1} y}\;, \psi_{2}(x,y)=\left(
       \begin{array}{c}
        {\bm 0} \\
        \mathcal{U}_{2}\\
       \end{array}
     \right)e^{ ik_{x}x}e^{\lambda_{2} y}\,,
\end{equation}
which gives the energy of the edge states for each block, $i=1,2$,
\begin{equation}\label{displinear}
  E_i=(-1)^{i+1} A_{2}k_{x}\cos(\vartheta)+|t(0,a)|\sin(\vartheta)\,,
\end{equation}
and the corresponding decay lengths
\begin{equation}
\label{lambda_linear}
\lambda_i=-\left\vert\frac{|t(0, a)|}{A_2}\cos\vartheta+ (-1)^i k_x\sin\vartheta\right\vert\,.
\end{equation}
Note that at the value of $k_x$ when $\lambda=0$, the edge states merge with the energy bands of the thin film, and therefore, indeed, are no longer localized.

The spinor components of the wave functions for these states are given by
\begin{equation}\label{us}
\mathcal{U}_{i}=\mathcal{N}_{i}\left(
                  \begin{array}{c}
                   i\eta[ {\Upsilon_i}(\sin\vartheta+\eta)-\mathcal{A}_{i}\cos\vartheta]\\
         -\eta[{\Upsilon_i}\cos\vartheta+\mathcal{A}_{i}(\sin\vartheta-\eta) ]\\
                  \end{array}
                \right)\,,
\end{equation}
where ${\Upsilon_i}=A_{2}(k_{x}-\lambda_i)$,  $\mathcal{A}_{i}=(E_i+\eta|t(0,a)|)$, $\eta=(-1)^{i+1}$ as before, and $\mathcal{N}_{i}$ is the normalization constant.

An explicit check shows that the states corresponding to the same energy and different indices $i$ are partners under time reversal, $\mathcal{T}=i\alpha_{x}\Sigma_{y}\mathcal{C}$, and are therefore stable with respect to TRS-invariant disorder. These counterpropagating edge states carry pseudospin
\begin{equation}\label{pseudospitextures}
\langle\alpha_{0}\Sigma_{x}\rangle_i=0\;,\langle\alpha_{0}\Sigma_{y}\rangle_i=\eta\cos\vartheta\,,\langle\alpha_{0}\Sigma_{z}\rangle_i=\eta\sin\vartheta\,.
\end{equation}
In the particle-hole symmetric case, $\vartheta=0$, when the only surviving pseudospin component is along $y$, the edge eigenstates become the eigenstates of the $\Sigma_y$ matrix and $E_{i}(k_x=0)=0$ as in the preceding section. When the particle-hole symmetry is broken, and the crossing point of the two branches shifts from $E=0$, the $z$-component of the pseudospin is also carried by the edge states.

Previous work~\cite{thingap2,thingap3} considered the particle-hole symmetric boundary conditions and argued that the existence of the edge states carrying pseudospin implies a quantum spin Hall effect in the topological phases of the thin films. We believe this statement to be incorrect. Indeed, recalling the physical spin operators, $S_{x,y}=\hbar\alpha_{x}\Sigma_{x,y}/2$, $S_{z}=\hbar\alpha_{0}\Sigma_{z}/2$, it is clear that any edge states of the form of Eq.~\eqref{edgestateslinear} carry no spin component in the $x$-$y$ plane. Only if we break the particle-hole symmetry at the boundary do the edge states acquire an out-of-plane component of the spin, $\langle\psi| S_{z}|\psi\rangle=\pm\sin(\vartheta)/2$. Hence the analog of the quantum spin Hall effect only occurs under non-trivial boundary conditions.

This is consistent with the analysis of the spin content of the edge states at other boundaries. If we consider the opposite boundary (half-space $y<0$ occupied by the thin film), the branches corresponding to the opposite blocks switch energy and pseudospin content. Similarly, the edge states at the boundaries parallel to the $x$-axis, have the same linear dispersion but have $\langle\Sigma_{x}\rangle=\pm\eta\cos\vartheta$, $\langle\alpha_0\Sigma_{y}\rangle=0$ and $\langle\alpha_{0}\Sigma_{z}\rangle=\pm\eta\sin\vartheta$, with $+/-$ indicating opposite edges. Therefore a state circulating around the periphery of the thin film changes the pseudospin orientation but always carries the single, $z$, component of the physical spin in the same direction.

We conclude that the observation of the spin-Hall effect in the TI thin film can be only achieved in the presence of particle-hole symmetry breaking edge potentials, and even in that case the spin component carried by the state may be very small, in stark contrast to the findings in Refs.~\onlinecite{thingap2,thingap3}. There is an additional aspect of these states that makes their experimental detection difficult. Recall that the decay constant of these states at $k_x=0$ is $\lambda_0\equiv(|t(0, a)|/A_2)\cos\vartheta$. Recall also that the tunneling gap $t(0, a)$ is exponentially small in the film thickness, so that, for example for Bi$_2$Se$_3$, for $a=15$\AA \ (film thickness $2a=3$nm, or about 3 quintuple layers (QL)), $|t(0,a)|=4$meV, which corresponds to the decay length, $\lambda_{0}^{-1}$, of at least 80nm (for $\vartheta=0$).  For the thickness of about 10 quintuple layers the edge states extend over 10-100$\mu$m from the boundary. Therefore these states are very delocalized, with the spectral weight distributed over a wide range away from the edge, making their detection with spectroscopic techniques very challenging. For the observation of these states using transport measurements we need to move away from the particle-hole symmetric case, towards $\vartheta=\pi/2$ when their spin content becomes substantial. However, because of the spatial extent of the edge states, to avoid back scattering across the sample and/or hybridization of the states of the opposite edges, one needs to have a good quality 3QL thick film with micron-size area, or a thicker film with an area of several square millimetres. This suggests that the observation of the edge state in Bi$_2$Se$_3$ thin films is difficult. From the inset of Fig.~\ref{fig2}(b), Bi$_2$Te$_3$ is a much more promising candidate material for observation of the topological effects in thin films as $t(0,a)$ remains on the order of several meV even for close to 100\AA \ thick films. For a $3$~QL thin film of  Bi$_2$Te$_3$ the decay length, $\lambda_{0}^{-1}\approx 10$ nm, while for a $5$ QL thick it is about $30$ nm.

\section{Conclusions }\label{conclusions}

We showed that a thin film of topological insulator material can host both trivial and topological phases that are controllable via the thickness of the film. To reach these conclusions we introduced a tunneling formalism that generically applies to thin-film based heterostructures, and used it for the free-standing case. We demonstrated that two (hitherto missed) technical aspects are critical for the correct analysis of this problem. First, we had to analyze a general I-TI-I junction, and only then obtain the free-standing film limit by setting the insulator gap to infinity. Second, we had to keep track of the dependence of the tunneling matrix element (or mass) on the in-plane momentum of the film, as this dependence is non-monotonic, and crucial for the topological properties.

For films of Bi$_2$Se$_3$ and Bi$_2$Te$_3$ the direct spectral gap at the $\Gamma$-point (in plane momentum $\bm k=0$) of the surface Brillouin Zone oscillates as a function of the thickness, in agreement with Refs.~\onlinecite{thingap2,thingap3,thingap1}. However, in contrast to previous work that utilized the small-$\bm k$ expansion to draw conclusions about the topology, we find that: a) the topological properties are entirely determined by the sign of the tunneling matrix element at $\bm k=0$; b) all the topological transitions occur with a gap closing (and the results in Refs.~\onlinecite{thingap2,thingap3,thingap1} suggesting otherwise are artefacts of the expansion); c) even when the gap closes and the film is a semimetal, the states at the opposite surfaces are coupled due to the $\bm k$-dependence of the tunneling. This coupling is manifested as the band curvature away from the Dirac point, and therefore can be tested in doped samples via conventional transport coefficients or via non-linear transport measurements in the undoped films.

We investigated the topological properties of the film by analytically computing the pseudospin textures and numerically evaluating the Chern number to arrive at a topological phase diagram as a function of the film thickness. We confirmed the topologically non-trivial nature of the phases by determining the spectra and the wave functions of the edge states. We showed that, while there are counter-circulating {\em pseudospin} currents associated with the edge states, these states carry physical spin current only when the particle-hole symmetry is broken by the edge boundaries. In that latter case there is a net circulation of the out-of-plane spin component, whose magnitude depends on the specifics of the symmetry-breaking at the boundary. These findings are again in contrast to previous work~\cite{thingap2,thingap3,thingap1}, which argued for a quantum spin Hall state associated with the in-plane spin component of the edge modes for the particle-hole symmetric case. We find the particle-hole symmetric situation to be more reminiscent of the valley Hall effect in graphene~\cite{valleyHallgrpahen}.

In the absence of particle-hole symmetry the edge-states carry a non-quantized $z$-component of the spin. Therefore, the quantum spin-Hall effect is absent in the its usual sense of quantized spin transport. Nonetheless, a setup combining spin selection via the metallic spin-Hall transport with measurements of charge conductivity using split-gate techniques, would still observe quantized charge $\sigma_{xy}$ that has been interpreted as evidence for the quantum spin-Hall conductivity~\cite{molenkamp}.

We also found that the spatial extent of the edge states is very large in Bi$_2$Se$_3$, making their observation by either spectroscopic or transport measurements difficult, and requiring large area samples to avoid back-scattering and hybridization of the modes at the opposite edges. We suggested that Bi$_2$Te$_3$ is a more promising material for the observation of the edge states, since the oscillatory behavior of the tunneling matrix element is more pronounced there, and the edge states are more localized.

We believe that our method and the results shown here will stimulate theoretical and experimental studies of the topological phases of the thin films, and their potential use in next generation functional devices.

\acknowledgments
This research was supported by NSF via grants No. DMR-1410741 and No. DMR-1151717.

\bibliography{referencesthinfilm}

\end{document}